\documentclass[12pt,preprint,numberedappendix,twocolappendix]{emulateapj}

\newcommand\bibinc{n}		

\usepackage{subeqnarray}
\usepackage{amsmath}
\usepackage{hyperref}
\usepackage{ulem}
\usepackage{stmaryrd}

\hypersetup{colorlinks=true,linkcolor=black,citecolor=blue,urlcolor=black}

\usepackage{lineno}

\usepackage{natbib}
\def\tt{  }

\newcommand{\Eq}[1]{Equation\,(\ref{#1})}

\newcommand{\Sec}[1]{Section~\ref{#1}}

\newcommand{\Fig}[1]{Figure~\ref{#1}}

\newcommand {\mesa} {{\ttfamily MESA}}

\setcitestyle{citesep={,}}
\begin{document}

\slugcomment{ApJ, 893:36, 2020}

\shorttitle{Re-inflation of warm and hot Jupiters}
\shortauthors{Komacek, Thorngren, Lopez, \& Ginzburg}

\title{Re-inflation of warm and hot Jupiters}
\author{Thaddeus D. Komacek$^{1}$, Daniel P. Thorngren$^2$, Eric D. Lopez$^{3,4}$, and Sivan Ginzburg$^{5}$ } \affil{$^1$Department of the Geophysical Sciences, The University of Chicago, Chicago, IL, 60637, USA  \\ $^2$Institute for Research on Exoplanets, Universit\'{e} de Montr\'{e}al, Montr\'{e}al, Qu\'{e}bec, H3T 1J4, Canada \\ $^3$NASA Goddard Space Flight Center, Greenbelt, MD 20771, USA  \\ $^4$GSFC Sellers Exoplanet Environments Collaboration, USA \\ $^5$Department of Astronomy, University of California at Berkeley, CA 94720-3411, USA  \\ \url{tkomacek@uchicago.edu} }
\begin{abstract}
Understanding the anomalous radii of many transiting hot gas giant planets is a fundamental problem of planetary science. Recent detections of re-inflated warm Jupiters orbiting post-main-sequence stars and the re-inflation of hot Jupiters while their host stars evolve on the main-sequence may help constrain models for the anomalous radii of hot Jupiters. In this work, we present evolution models studying the re-inflation of gas giants to determine how varying the depth and intensity of deposited heating affects both main-sequence re-inflation of hot Jupiters and post-main-sequence re-inflation of warm Jupiters. We find that deeper heating is required to re-inflate hot Jupiters than is needed to suppress their cooling, and that the timescale of re-inflation decreases with increasing heating rate and depth. We find a strong degeneracy between heating rate and depth, with either strong shallow heating or weak deep heating providing an explanation for main-sequence re-inflation of hot Jupiters. This degeneracy between heating rate and depth can be broken in the case of post-main-sequence re-inflation of warm Jupiters, as the inflation must be rapid to occur within post-main-sequence evolution timescales. We also show that the dependence of heating rate on incident stellar flux inferred from the sample of hot Jupiters can explain re-inflation of both warm and hot Jupiters. TESS will obtain a large sample of warm Jupiters orbiting post-main-sequence stars, which will help to constrain the mechanism(s) causing the anomalous radii of gas giant planets.
\end{abstract}
\section{Introduction}
\indent The observation that many transiting hot Jupiters have radii larger than expected from standard evolutionary models is {an outstanding question of} exoplanetary science \citep{Guillot_2002,fortney_2009,Baraffe:2010xe,Baraffe:2014,Laughlin:2015,Laughlin:2018aa}. A variety of mechanisms have been propsed to explain the anomalous transit radii of hot Jupiters \citep{Weiss:2013,Baraffe:2014}, including tidal mechanisms \citep{Bodenheimer:2001,Gu:2003aa,Gu:2004aa,Jackson:681,Ibgui:2009,Miller:2009,Arras:2010,Ibgui:2010,Leconte:2010a,Gu:2019aa}, modifications to the microphysics of hot Jupiters \citep{Burrows:2007bs,Chabrier:2007,Leconte:2012,Kurokawa:2015aa}, incident stellar-flux-driven hydrodynamic mechanisms \citep{showman_2002,Guillot_2002,Youdin_2010,Tremblin:2017,Sainsbury-Martinez:2019aa}, and Ohmic dissipation \citep{Batygin_2010,Perna_2010_2,Batygin_2011,Huang_2012,Menou:2012fu,Rauscher_2013,Wu:2013,Rogers:2020,Rogers:2014,Ginzburg:2015a}. Studies of the radius distribution of hot Jupiters \citep{Demory:2011,Laughlin_2011,Miller:2011,Thorngren:2017} have shown that radius anomalies only occur for gas giants with equilibrium temperatures in excess of $1000~\mathrm{K}$. Additionally, \cite{Laughlin_2011,Weiss:2013}, and \cite{Thorngren:2017} showed that the radii of hot Jupiters correlate with incident flux. As a result, the mechanism that inflates hot Jupiters is directly tied to the incident flux from the host star. Recently, \cite{Thorngren:2017} found that the fraction of irradiation that is converted to deposited heat must peak at an intermediate equilibrium temperature of $\sim 1600~\mathrm{K}$ and fall off for both hotter and colder planets.   \\
\indent \cite{Lopez:2015} predicted that warm Jupiters {will re-inflate if their equilibrium temperature crosses the $1000~\mathrm{K}$ heating threshold as their host stars evolve, provided that sufficient heat is deposited deep within the planet.}
Recent K2 observations of warm Jupiters orbiting post-main-sequence stars have found three candidate re-inflated planets \citep{Grunblatt2016,Grunblatt:2017aa,Grunblatt:2019aa}. All of these planets have significantly inflated radii of $\approx 1.3-1.45~R_\mathrm{Jup}$, which can be explained by heating at the very center of the planet with a deposited heating rate that is $\approx 0.03\%$ of the incident stellar power \citep{Grunblatt:2017aa}. \cite{Hartman2016} found evidence that hot Jupiters re-inflate while their host stars brighten during main-sequence evolution. Main-sequence re-inflation requires deposited heat, as mechanisms that only slow interior cooling cannot cause an increase in the planetary radius over time. \cite{Thorngren-et-al.:2019aa} confirmed the finding of main-sequence re-inflation using a Bayesian structural analysis of 232 hot Jupiters, finding evidence for a correlation between planetary radius and fractional age (age normalized by the main-sequence lifetime) of the host star. Additionally, \cite{Thorngren-et-al.:2019aa} found that the radii of hot Jupiters track the incident flux from their host stars, not the age of the star. These observations of both main-sequence and post-main-sequence re-inflation show that the incident stellar flux and deposited heating rates are linked --- if the deposited heating rate were constant, the radius of the planet would either decrease or stay constant over time.  \\
\indent \cite{Lopez:2015} found that re-inflation can occur in the limiting case of heat that is deposited at the center of the planet. However, studies using heating profiles relevant for individual dissipation mechanisms differ on whether shallower heating can re-inflate hot Jupiters. \cite{Batygin_2011} found that Ohmic dissipation can cause re-inflation of hot Jupiters, while \cite{Wu:2013} and \cite{Ginzburg:2015a} found that Ohmic dissipation can only stall contraction, not lead to significant re-inflation. This is because heating re-inflates planets from the heating level downward, and the timescale for deposited heating to warm up the interior of the planet scales inversely with the heating depth \citep{Ginzburg:2015a}. For Ohmic dissipation, \cite{Ginzburg:2015a} found that the timescale for re-inflation is $\sim 30~\mathrm{Gyr}$, much longer than the $\sim 1~\mathrm{Gyr}$ cooling timescale. A key difference between the numerical models of \cite{Batygin_2011} and \cite{Wu:2013} is that \cite{Batygin_2011} include the increase in incident stellar power with increasing planetary radius while \cite{Wu:2013} do not. For a fixed conversion rate of incident stellar power to energy deposition, this leads to an increase in the deposited heat with increasing planetary radius. In this paper, we will show that including this feedback between the planetary radius, incident stellar power, and heating rate can enhance re-inflated planet radii. \\
\indent The constraints derived by \cite{Thorngren:2017}  on the heating rate needed to explain the sample of inflated hot Jupiters assume that the heat is deposited at the very center of the planet. However, \cite{Spiegel:2013,Ginzburg:2015}, and \cite{Komacek:2017a} showed that there is a degeneracy between the heating rate and the depth of heating --- deeper heating requires weaker heating rates to lead to a given radius, and vice versa. Observations of re-inflated warm Jupiters orbiting post-main-sequence stars provide an avenue in which this degeneracy can be broken. This is because the re-inflation timescale is strongly dependent on the depth of heating \citep{Ginzburg:2015a}, {which is dependent on the heating mechanism}. Because post-main-sequence evolution timescales are fast ($\sim 100~\mathrm{Myr}$), only sufficiently deep heating will lead to post-main-sequence re-inflation of warm Jupiters.    \\
\indent In this paper, we study both the re-inflation of hot Jupiters while their host stars evolve on the main-sequence (which we term ``main-sequence re-inflation'') and the re-inflation of warm Jupiters while their host stars evolve on the post-main-sequence (termed ``post-main-sequence re-inflation''). This work builds off of that of \cite{Lopez:2015}, and uses a similar methodology to \cite{Komacek:2017a}. We improve on previous work by studying how varying both heating rate and depth affect re-inflation. Additionally, we study both main-sequence and post-main-sequence re-inflation with a unified framework. Lastly, we show how the degeneracy between heating rate and heating depth can be broken with future observations of re-inflated gas giants. \\
\indent This paper is organized as follows. In \Sec{sec:methods}, we describe our model setup and each of our three simulation grids studying re-inflation. The results of these numerical experiments are shown in \Sec{sec:results}. We develop analytic theory for re-inflation due to point source energy deposition in \Sec{sec:theory}, and compare our theory to the results of our numerical experiments. We discuss our results and describe how observations of re-inflation can test inflation mechanisms in \Sec{sec:disc}, and conclude in \Sec{sec:conc}. 

\section{Methods}
\label{sec:methods}
\subsection{Numerical model}
In this work, we use the {\mesa}\ stellar and planetary evolution code \citep{Paxton:2011,Paxton:2013,Paxton:2015,Paxton:2018aa,Paxton:2019aa} to solve the time-dependent equations of stellar structure \citep{Chandrasekhar:1939,Kippenhahn:2012} applied to gas giant planets. {Our modeling framework is one-dimensional (1D), and as a result does not take into account either changes in the planetary structure as a function of latitude and longitude or atmospheric dynamics that can act to transport heat, limitations that are both described in further detail below. The planetary structure equations we solve include} the mass conservation equation, 
\begin{equation}
\label{eq:mass}
\frac{dm}{dr} = 4 \pi r^2 \rho \mathrm{,}
\end{equation}
where $m$ is the enclosed mass at a radius $r$, with mass density $\rho$. We ensure hydrostatic equilibrium,
\begin{equation}
\frac{dP}{dm} = -\frac{Gm}{4 \pi r^4} \mathrm{,}
\end{equation}
where $P$ is the pressure and $G$ is the gravitational constant. Energy conservation is included as
\begin{equation}
\frac{dL}{dm} = \epsilon_{\mathrm{grav}} + \epsilon_\mathrm{irr} + \epsilon_\mathrm{dep}  \mathrm{,}
\end{equation}
where $L$ is the outgoing luminosity, $\epsilon_{\mathrm{grav}} = -T dS/dt$ (where $T$ is temperature) is the {loss or gain of entropy ($S$) due to gravitational contraction or inflation}, {$\epsilon_\mathrm{irr}$ is additional heating due to irradiation, and $\epsilon_\mathrm{dep}$ represents internal heat deposition. We describe our choices for $\epsilon_\mathrm{irr}$ and $\epsilon_\mathrm{dep}$ in further detail below}. Lastly, we solve the energy transport equation,
\begin{equation}
\label{eq:energytrans}
\frac{dT}{dm} = -\frac{G m T}{4 \pi r^4 P}\nabla \mathrm{,}
\end{equation}  
where $\nabla \equiv d \mathrm{ln} T / d \mathrm{ln} P$ is the logarithmic temperature gradient, set equal to the smaller of the adiabatic gradient $\nabla_\mathrm{ad}$ or radiative gradient $\nabla_\mathrm{rad}$. 
In {radiative} regions, the temperature gradient is set equal to the radiative gradient 
\begin{equation}
\nabla_\mathrm{rad} = \frac{3}{64 \pi \sigma G}\frac{\kappa L P}{m T^4} \mathrm{,}
\end{equation} 
where $\sigma$ is the Stefan-Boltzmann constant and $\kappa$ is the opacity, updated from \cite{Freedman:2008} as described in \cite{Paxton:2013} and assuming a dust-free Solar composition. {We use a zero-width radiative-convective boundary and do not model convective overshoot, which would cause the exchange of energy in both directions across the radiative-convective boundary \citep{Youdin_2010,Leconte:2012}. Additionally, our use of a 1D modeling framework does not consider the possibility that the radiative-convective boundary is non-uniform \citep{Budaj:2012aa,rauscher_showman_2013}.}  Equations (\ref{eq:mass})-(\ref{eq:energytrans}) are closed using the \mesa \ equation of state \citep{Paxton:2019aa}, which is largely from \cite{Saumon:1995} for the temperatures and densities relevant for gas giant planets. \\
\indent We use the same basic model setup as \cite{Komacek:2017a}, studying gas giants that are both externally irradiated and have deposited heating in their atmospheres or interiors. However, instead of studying how heating slows the radius contraction of hot Jupiters as in \cite{Komacek:2017a}, in this work we study the re-inflation of both warm and hot Jupiters. We study re-inflation using three separate model grids: an idealized suite studying the process by which re-inflation occurs, a suite of models studying re-inflation of hot Jupiters during main-sequence evolution of their host stars, and a suite studying the re-inflation of warm Jupiters orbiting post-main-sequence stars. These model grids are described in detail in \Sec{sec:grids}.  \\
\indent We incorporate irradiation and deposited heating by adding extra energy terms {$\epsilon_\mathrm{irr}$ and $\epsilon_\mathrm{dep}$} to the energy conservation equation, as in \cite{Komacek:2017a}. The incoming stellar flux $F_\star$ is incorporated as an energy generation rate
\begin{equation}
\epsilon_\mathrm{irr} = \frac{F_\star}{4\Sigma_p} \mathrm{,}
\end{equation} 
applied in an outer mass column $\Sigma_p$ of the planet as in \cite{Valsecchi:2015,Owen:2015}, and \cite{Komacek:2017a}. We describe our choices for $\Sigma_p$ in the following \Sec{sec:grids}. This irradiation leads to a slight increase in the radius relative to non-irradiated models, but {when implemented in 1D structure models} cannot explain the radius inflation of many hot Jupiters \citep{Arras:2006kl,Fortney:2007ta}. {Irradiation powers atmospheric circulation that acts to transport heat both from day-to-night \citep{Perez-Becker:2013fv,Komacek:2015,Komacek:2017} and vertically \citep{Youdin_2010,Tremblin:2017,Zhang:2018aa,Komacek:2019aa,Sainsbury-Martinez:2019aa}, but this is not included in our modeling framework.} \\
\indent We model deposited heating as an additional term in the extra energy dissipation rate $\epsilon_\mathrm{extra}$, as was done in previous studies of gaseous planet evolution with {\mesa} \citep{Wu:2013,Komacek:2017a,Millholland:2019aa}. {This framework models direct heat deposition, and does not take into account heat transport by, e.g., the deep atmospheric circulation \citep{Sainsbury-Martinez:2019aa}.} The heating rate $\epsilon_\mathrm{dep}$ is set to be a Gaussian in pressure with a standard deviation of half of a pressure scale height, as in \cite{Komacek:2017a}.
We consider a range of integrated heating rates
\begin{equation}
\Gamma = \int_0^{M_p} \epsilon_\mathrm{dep} dm \mathrm{,}
\end{equation}
where $M_p$ is the mass of the planet. We set the integrated heating rates to different fractions of the incident stellar power as
\begin{equation} 
\gamma = \frac{\Gamma}{L_\mathrm{irr}} \mathrm{,}
\end{equation} 
where the incident stellar power is 
\begin{equation}
L_\mathrm{irr} = \pi R_p^2 F_\star \mathrm{,}
\end{equation}
with $R_p$ the radius of the planet at the photosphere, where the optical depth to incoming radiation $\tau = 2/3$. We vary $\gamma$ between $10^{-5}$ and $0.1$ in all of our simulation grids. We consider heating centered at deposition pressures $P_\mathrm{dep}$ ranging from $1~\mathrm{bar}$ to $10^6~\mathrm{bars}$, and include cases with heating at the very center of the planet. \\
\indent For all of our simulations, we use an initial model of {an HD 209458b analogue} with a mass of $0.69~M_\mathrm{Jup}$, a composition with a helium fraction $Y = 0.24$, metallicity $Z = 0.02$, and without a heavy element core as in the HD 209458b models of \cite{Guillot_2002} and \cite{Komacek:2017a}. The stopping points of our simulations are different for each model grid, as described in the following \Sec{sec:grids}.

\subsection{Simulation grids}
\label{sec:grids}
We conduct three separate grids of {\mesa} simulations to study the re-inflation of gas giants, as described below. 
\subsubsection{Re-inflation of an evolved hot Jupiter}
Our first suite of models studies the re-inflation of an evolved hot Jupiter that undergoes fixed rates of irradiation and deposited heating. These simulations are idealized and do not directly apply to either the case of main-sequence re-inflation of hot Jupiters or post-main-sequence re-inflation of warm Jupiters. However, they are useful to understand the process by which planets re-inflate, and we compare the results from this suite of numerical experiments to analytic theory in \Sec{sec:theory}. The starting point for these simulations is an HD 209458b model which has been evolved for $10~\mathrm{Gyr}$ without any deposited heating, with a final radius of $1.08 R_\mathrm{Jup}$. We then re-inflate the planet for $10~\mathrm{Gyr}$ including deposited heating with varying heating rate and depth. \\
\indent In this suite of simulations, we keep the incident stellar flux fixed at $F_\star = 1.0012 \times 10^9~\mathrm{erg}~\mathrm{cm}^{-2}~\mathrm{s}^{-1}$, which corresponds to a full-redistribution equilibrium temperature of $T_\mathrm{eq} = 1450~\mathrm{K}$. The outer mass column in which irradiation is applied is also fixed at $\Sigma_p = 250~\mathrm{g}~\mathrm{cm}^{-2}$. Our chosen $\Sigma_p$ is equal to a visible opacity of $\kappa_\mathrm{vis} = 4 \times 10^{-3} \mathrm{cm}^2~\mathrm{g}^{-1}$, as used in \cite{fortney-etal-2008,Guillot:2010}, and \cite{Owen:2015}. {For this visible opacity, the $\tau = 1$ level to incoming irradiation lies at a pressure of $0.23~\mathrm{bars}$ for the present-day radius of HD 209458b.} These values of incident stellar flux and irradiated column mass are the same as used in \cite{Komacek:2017a}. Additionally, in this suite of simulations we keep the heating rate fixed in time, and do not include the increase in the heating rate due to the increasing planetary cross-sectional area. Instead, as in \cite{Komacek:2017a} the heating rate is kept to a fixed fraction of the present-day incident stellar power of HD 209458b, which is $2.4 \times 10^{29}~\mathrm{erg}~\mathrm{s}^{-1}$. This model suite can hence be considered as the planetary re-heating analogue to the simulations of \cite{Komacek:2017a} that studied how heating can slow planetary cooling. We describe the results from this simulation grid and directly compare to the results of \cite{Komacek:2017a} in \Sec{sec:idealized}.  

\subsubsection{Main-sequence re-inflation}
In our second suite of simulations, we model how the evolution of a hot Jupiter undergoing deposited heating is affected by the varying luminosity of the host star. To do so, we incorporate a time-dependent incident stellar flux $F_\star = L_\star/(4\pi a^2)$ using pre-calculated stellar evolution tracks from MIST models \citep{Choi:2016aa,Dotter:2016aa} to obtain the stellar luminosity $L_\star$. We assume a fixed planetary semi-major axis of $a = 0.04747~\mathrm{au}$ relevant for HD 209458b. We include deposited heating in the planet throughout the main-sequence evolution of its host star, keeping the fraction of the incident stellar power converted to deposited heating ($\gamma$) fixed with time. {Note that though we keep $\gamma$ fixed in our main grid of simulations, in \Sec{sec:test} we include the inferred dependence of deposited heating on equilibrium temperature from \cite{Thorngren:2017} in our evolution models}. We stop these models when the host star reaches the end of the main-sequence, which occurs at $9.88~\mathrm{Gyr}$ for our simulations of planets orbiting a Sun-like star. \\
\indent In both this suite of simulations and the suite studying post-main-sequence re-inflation (described in \Sec{sec:pmsdescribe}), we keep the outer mass column in which irradiation is applied fixed at $\Sigma_p = 300~\mathrm{g}~\mathrm{cm}^{-2}$. This corresponds to a visible opacity of $\kappa_\mathrm{vis} = 3.33 \times 10^{-3} \mathrm{cm}^2~\mathrm{g}^{-1}$ {and a visible photosphere at $0.27~\mathrm{bars}$ when the radius is equal to that of HD 209458b}. We use this reduced visible opacity to aid with model stability at times in the host star evolution when the incident stellar flux rapidly increases. We show results from our main-sequence re-inflation grid in \Sec{sec:mainseq}. 

\subsubsection{Post-main-sequence re-inflation}
\label{sec:pmsdescribe}
Our third grid of simulations studies the evolution of warm Jupiters that re-inflate while their host star evolves on the post-main-sequence. In this suite of models, we only include deposited heating if the incident stellar flux $F_\star \ge 2.268 \times 10^8~\mathrm{erg}~\mathrm{cm}^{-2}~\mathrm{s}^{-1}$, which corresponds to an equilibrium temperature $T_\mathrm{eq} \ge 1000~\mathrm{K}$. We do so because gas giants with $T_\mathrm{eq} < 1000~\mathrm{K}$ do not have anomalous radii \citep{Demory:2011,Laughlin_2011,Miller:2011,Lopez:2015,Thorngren:2017}. Weak deposited heating in warm Jupiter interiors is also expected from the inferred dependence of deposited power on $T_\mathrm{eq}$ \citep{Thorngren:2017}, which decreases to zero at $T_\mathrm{eq} < 1000~\mathrm{K}$. This is also consistent with Ohmic dissipation {and models of atmospheric heat transport}, which expect that planets with $T_\mathrm{eq} < 1000~\mathrm{K}$ should {not be inflated due to the small day-night forcing and low atmospheric ionization fraction \citep{Youdin_2010,Menou:2012fu,Ginzburg:2015a,Tremblin:2017}}. As a result, we assume that there is no deposited heating for planets with $T_\mathrm{eq} < 1000~\mathrm{K}$, because otherwise warm Jupiters with anomalously large radii would have been discovered. To support this assumption, we show in \Sec{sec:mainseqdisc} that if gas giants with $T_\mathrm{eq} < 1000~\mathrm{K}$ did undergo deposited heating with a similar conversion rate of incident stellar power to deposited heat and heating depth as inflated hot Jupiters, warm Jupiters would likely be inflated as well.  \\
\indent In our post-main-sequence re-inflation simulations, we study planets that lie at equilibrium temperatures below $1000~\mathrm{K}$ for the majority of the time that their host stars are on the main-sequence. As a result, the inflation mechanism heats the planet only after the host star is at or near the end of its main-sequence evolution. Our fiducial case is that of a warm Jupiter with an orbital separation of $0.1~\mathrm{au}$ orbiting a Sun-like star, which corresponds to an equilibrium temperature of $882~\mathrm{K}$ for the present-day Solar luminosity.  We use the same stellar evolution tracks as for our main-sequence re-inflation models, but evolve our simulations until the host star reaches a radius of $10~R_\varodot$. This corresponds to an age of $11.27~\mathrm{Gyr}$ for a planet orbiting a Sun-like star, which occurs while the star is on the red giant branch. We choose this stopping radius because it is challenging to detect Jupiter-sized planets around larger stars with current instrumentation \citep{Lopez:2015}, {and because after this point the radius of the host star quickly grows and the planet would become engulfed}. Results from these simulations studying post-main-sequence re-inflation are shown in \Sec{sec:pms}.

\section{Results}
\label{sec:results}
\subsection{Re-inflation of an evolved hot Jupiter}
\label{sec:idealized}
\begin{figure}
\includegraphics[width=0.5\textwidth]{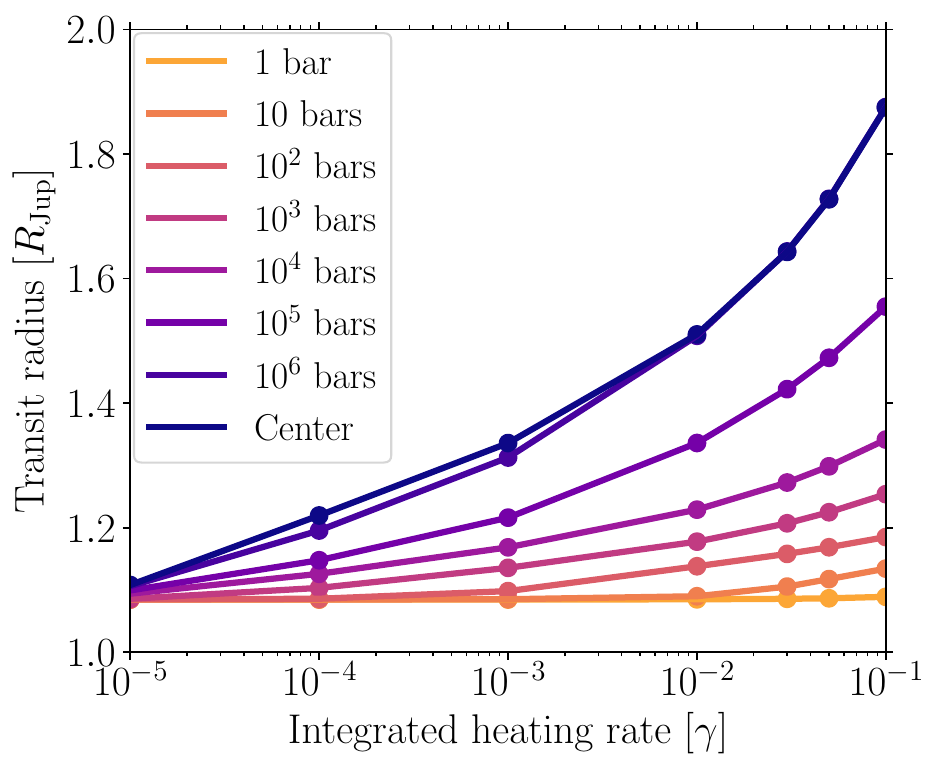}
\caption{\textbf{Planets that undergo deep heating can significantly re-inflate.} Shown is the transit radius in Jupiter radii after $10~\mathrm{Gyr}$ of re-heating for varying integrated heating rates ($\gamma = \Gamma/L_\mathrm{irr}$, from $10^{-5}$ to $0.1$) and heating locations ($P_\mathrm{dep}$, from $1~\mathrm{bar}$ to the planet center, with darker colors corresponding to deeper heating). Planets have the mass of HD 209458b and receive a fixed irradiation power of $2.4 \times 10^{29}~\mathrm{erg}~\mathrm{s}^{-1}$. We find that heating that is stronger and/or deeper leads to greater re-inflation.}
\label{fig:radii}
\end{figure}
\indent To elucidate the process by which gas giants re-inflate, we first analyze the results from our suite of idealized simulations of the re-inflation of an evolved hot Jupiter. \Fig{fig:radii} shows the transit radius after $10~\mathrm{Gyr}$ of re-heating for a hot Jupiter with an initial radius of $1.08~R_\mathrm{Jup}$ for varying heating rates $\gamma$ and heating depths $P_\mathrm{dep}$. Note that the pressure level of heating at the center of the planet depends on the heating rate, varying from $12.1$ Mbar with a weak heating rate of $\gamma = 10^{-5}$ after 10 Gyr of re-heating to $4.35$ Mbar with a strong heating rate of $\gamma = 10^{-1}$ at the same age. To calculate the transit radius from the photospheric radius, we use the isothermal limit of \cite{Guillot:2010} (see their Equation 60) and set the ratio of visible to infrared opacities equal to 0.4, as in \cite{Komacek:2017a}. We find that the transit radius increases monotonically with both integrated heating rate and heating depth. As a result, increasing either the heating rate or the heating depth leads to greater re-inflation. We find that deep heating at or near the center of the planet can lead to significant re-inflation, as in \cite{Lopez:2015}. \\
\indent Comparing our results in \Fig{fig:radii} for the effect of deposited heat on re-inflation to the effect of deposited heat on slowing planetary cooling from Figure 3 of \cite{Komacek:2017a}, we find significant differences. For re-inflation, there is not a large increase in the transit radius between 10 and 100 bars and the radius continues to increase with deeper heating within the interior (at pressures $P_\mathrm{dep} \ge 10^3~\mathrm{bars}$), unlike that found in \cite{Komacek:2017a}. This shows that, at a given age, the effects of deposited heating on re-inflation are fundamentally different than the effects of heating on offsetting the cooling of an initially high-entropy planet. However, we will show in \Sec{sec:theory} that the final equilibrium state (at a time $t = \infty$) of planets that undergo heating which leads to re-inflation and that undergo heating which delays planetary cooling is the same. 
\begin{figure}
\includegraphics[width=0.5\textwidth]{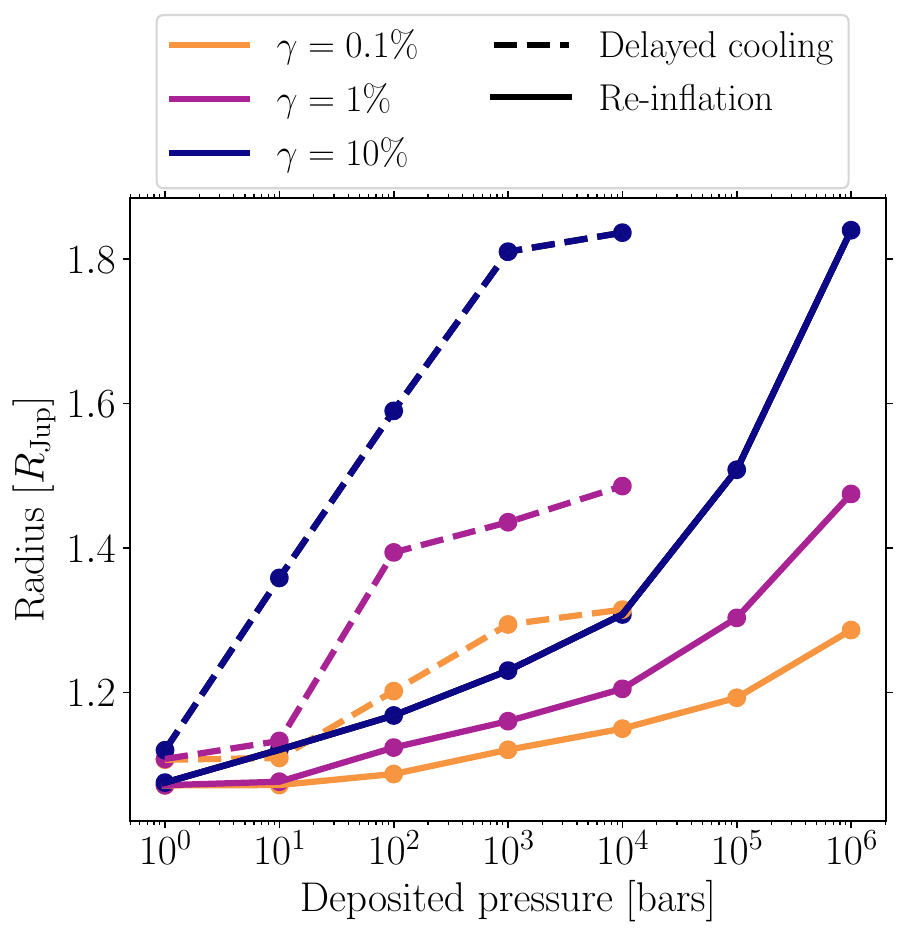}
\caption{\textbf{Heating needs to be deeper to re-inflate planets than it does to delay planetary cooling.} Shown is a comparison of our results for how re-inflated planet radii after $5~\mathrm{Gyr}$ of evolution depend on deposited pressure for varied integrated heating rates $\gamma$ (solid lines) with the results of \cite{Komacek:2017a} (dashed lines), who considered the effect of heating on delaying planetary cooling. We find that heating that leads to re-inflation has to be at pressures of $10^6~\mathrm{bars}$ or greater to reach a similar radius to that in delayed cooling models with heating deeper than $10^3~\mathrm{bars}$.}
\label{fig:radii_delayvsinf}
\end{figure}
\\ \indent \Fig{fig:radii_delayvsinf} directly compares our results for the effect of heating on re-inflation and the results of \cite{Komacek:2017a} on the effect of heating on slowing planetary cooling. {We find that re-inflation requires heating at pressures $P_\mathrm{dep} \ge 10^6~\mathrm{bars}$ to reach the same radius at $5~\mathrm{Gyr}$ as delayed cooling models with heating at pressures $P_\mathrm{dep} \ge 10^3~\mathrm{bars}$}. Unlike deposited heating that delays planetary cooling, the radii of re-inflated planets after $5~\mathrm{Gyr}$ continue to increase with deeper heating deposited below the inner radiative-convective boundary. Deposited heating that leads to re-inflation heats the {planet both upward and downward of the deposition level. Re-inflation from the heating level upward (which we term ``inside-out'' re-inflation) occurs very quickly, within $\lesssim 1~\mathrm{Myr}$ in most cases. Meanwhile, the timescale to re-heat the center (termed ``outside-in'' re-inflation) can be as long as Tyrs} and decreases with increasing depth of heat deposition \citep{Ginzburg:2015a}. {We explore the differences between inside-out and outside-in re-inflation in detail in Appendix \ref{sec:appendixa}.} \\
\indent {Because the re-inflation timescale scales inversely with the heating depth}, deeper heating will lead to greater re-inflation, unlike in the case of delayed cooling where deposited heating below the inner radiative-convective boundary (at $P_\mathrm{dep} > 10^3~\mathrm{bars}$) leads to similar radii after $5~\mathrm{Gyr}$ of evolution \citep{Komacek:2017a}. This is because heating that slows planetary cooling only has to {balance} cooling from the interior convective zone. {Meanwhile, heating that re-inflates an initially cold planet has to increase the entropy at the center of the planet rather than simply reduce the internal cooling rate}. As long as it is deposited below the inner radiative-convective boundary, heating that acts to slow planetary cooling has almost the same effect on evolution regardless of deposition pressure, {while the radius after re-heating of an initially cold planet continues to increase with deeper heating within the internal convective zone}. 
\begin{figure}
\includegraphics[width=0.5\textwidth]{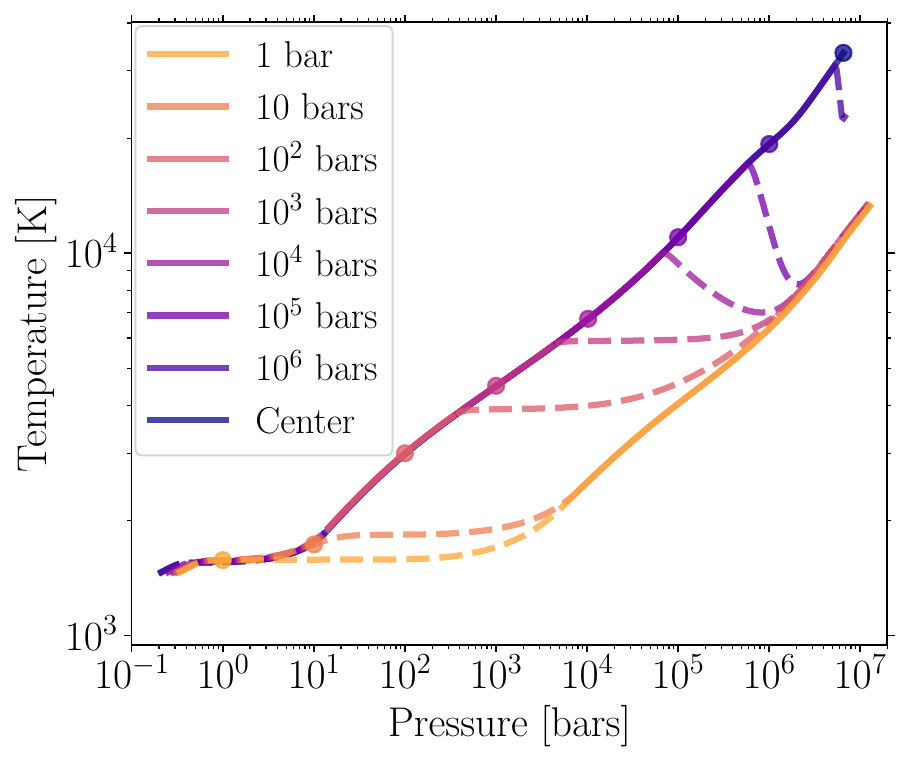}
\caption{\textbf{Deposited heating re-inflates the interior of a planet from the outside-in.} Temperature-pressure profiles from simulations after $10~\mathrm{Gyr}$ of re-heating with an integrated heating rate of $\gamma = 1\%$ and varying deposition pressure from $1~\mathrm{bar}$ to the planet center. Solid lines show convective regions, while dashed lines show {non-convective} regions. {Note that regions near the surface are radiative down to $\approx 10~\mathrm{bars}$}. Circles show the maximal heating location, with the color of the circle the same as the matching temperature-pressure profile. 
The entire interior of the planet is convective only in the case with heating at the very center.}
\label{fig:TP}
\end{figure}
\\ \indent 
\Fig{fig:TP} shows temperature-pressure profiles from simulations with a fixed heating rate of $\gamma = 1\%$ of the incident stellar power and varying heating depth. These temperature-pressure profiles are similar to those expected from the re-inflation models of \cite{Wu:2013} (see their Figure 7) and theory of \cite{Ginzburg:2015a} (see their Figure 5). {However, there are} differences due to our use of localized heat deposition instead of the Ohmic dissipation heating profiles considered in \cite{Wu:2013} and \cite{Ginzburg:2015a}, {and generally different heating mechanisms will lead to significant differences in the temperature profile}. We find that in the case of re-inflation, heating forces regions at pressures less than $P_\mathrm{dep}$ to be convective, similar to the case of heating that slows planetary cooling \citep{Komacek:2017a}. However, as in \cite{Wu:2013}, we find that heating that leads to re-inflation forces a downward heat flux that acts to re-inflate the planet from the heating level downward. As a result, the re-inflation timescale is governed by the downward {heat} flux from the heating level. We stress that the cases shown in \Fig{fig:TP} with $P_\mathrm{dep} \le 10^5~\mathrm{bars}$ are still evolving, while the final equilibrium (discussed in \Sec{sec:theory}) is characterized by an isotherm from the heating level to the center of the planet. We find that deposited heating that is not near the center has a relatively small effect on the central temperature and hence entropy of the internal adiabat after $10~\mathrm{Gyr}$ of evolution. As a result, only heating near the center can lead to re-inflation that greatly increases the radius of the planet over short timescales. 
\begin{figure}
\includegraphics[width=0.5\textwidth]{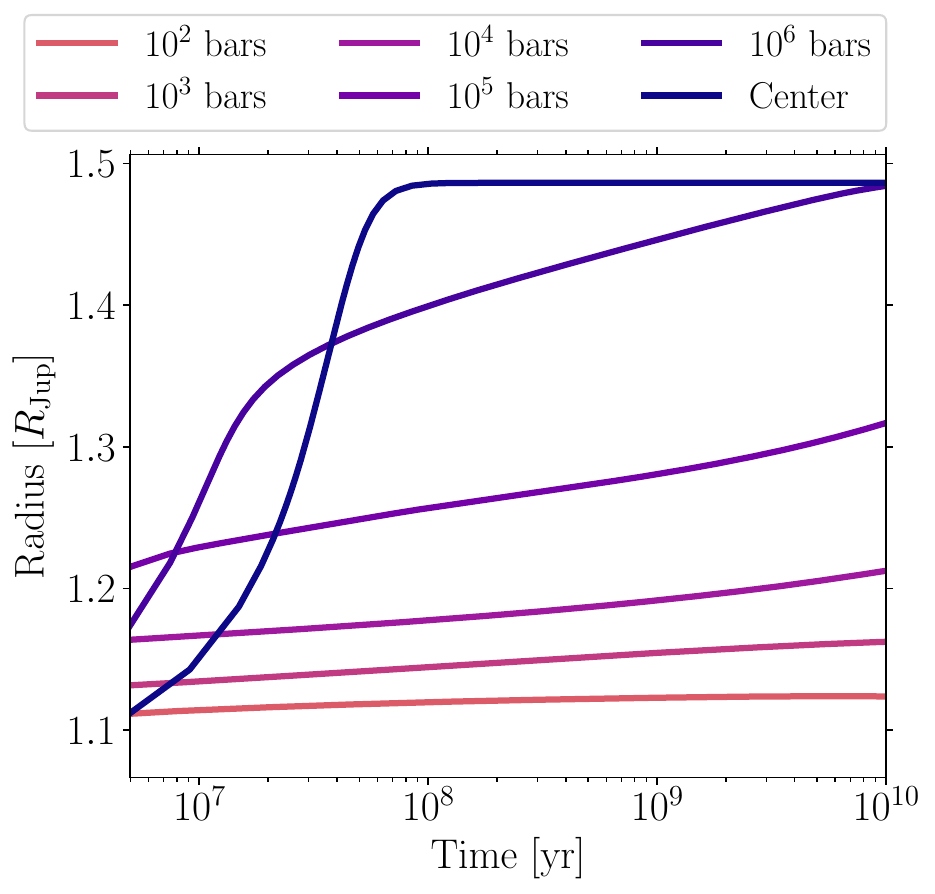}
\caption{\textbf{Re-inflation takes billions of years if heating is not deposited at the very center of the planet.} Radius evolution for simulations with a fixed heating rate of $\gamma = 1\%$, fixed incident stellar power of $L_\mathrm{irr} = 2.4 \times 10^{29}~\mathrm{erg}~\mathrm{s}^{-1}$, and varying heating locations from $100~\mathrm{bars}$ to the planet center. Only the simulation with heating at the very center reaches a steady state, while shallower heating models inflate over their evolution. The simulation with heating at $P = 10^6~\mathrm{bars}$ reaches a similar radius to the case with heating at the very center, while simulations with shallower heating reach smaller radii after re-inflation. \Fig{fig:rad_tp_longruns} shows the radius evolution in an extension of these simulations to $10^{13}$ yr, by which point simulations with $\gamma = 1\%$ and $P_\mathrm{dep} > 10^2~\mathrm{bars}$ have reached a radius equilibrium.}
\label{fig:rad_age}
\end{figure}
\\ \indent We find from our simulations that the timescale to re-inflate a planet decreases with increasing heating depth. \Fig{fig:rad_age} shows the radius evolution of simulations with fixed $\gamma = 1\%$ of the incident stellar power and varying heating depth. We find that the re-inflation timescale for heating at the center of the planet is $\lesssim 50~\mathrm{Myr}$, comparable to the initial cooling timescale before heating acts to slow planetary cooling (before regime 2 of \citealp{Komacek:2017a}). Deep heating at $10^6~\mathrm{bars}$ that is near the center can re-inflate planets to the same radius as central heating, but it requires billions of years over which the planet can re-inflate. Meanwhile, shallow heating at pressures $<10^3~\mathrm{bars}$ does not greatly affect the radius even after $10~\mathrm{Gyr}$ of evolution. {We will show in \Sec{sec:theory} that the long evolutionary timescales for planets with shallow heating are the cause of the differences in the dependence of radius on heating depth for re-inflation relative to delayed cooling shown in \Fig{fig:radii_delayvsinf}.} To summarize, we expect that shallow heating at pressures $\lesssim 10^2~\mathrm{bars}$ will not lead to re-inflation, moderately deep heating at pressures $10^3 \lesssim P_\mathrm{dep} \lesssim 10^5~\mathrm{bars}$ will lead to moderate re-inflation, and deep heating at pressures $\gtrsim 10^6~\mathrm{bars}$ will greatly re-inflate planets. 
\subsection{Main-sequence re-inflation}
\label{sec:mainseq}
\begin{figure}
\includegraphics[width=0.5\textwidth]{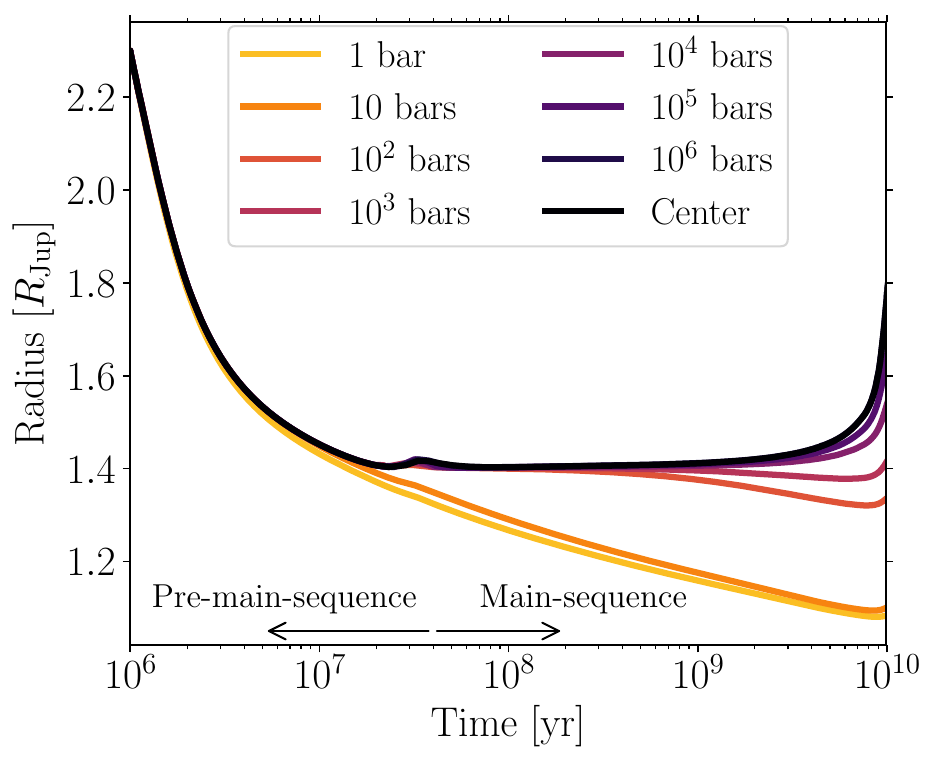}
\caption{\textbf{The radii of hot Jupiters that undergo deep heating evolve along with their host stars.} Radius evolution for simulated hot Jupiters orbiting a star with the stellar evolution track of the Sun and a fixed heating rate of $\gamma = 1\%$ for varying heating locations from $1~\mathrm{bar}$ to the planet center. The distinction between the pre-main-sequence and main-sequence phase of stellar evolution is shown by the arrows. Simulated planets have the mass and semi-major axis of HD 209458b. Heating must be deeper than $100~\mathrm{bars}$ to lead to re-inflation over the stellar main-sequence, and deeper heating leads to larger inflation over the stellar main-sequence lifetime.}
\label{fig:rad_age_hot}
\end{figure}
\indent Now we analyze the results from our suite of models studying the main-sequence re-inflation of hot Jupiters. \Fig{fig:rad_age_hot} shows radius evolution tracks for simulations with $\gamma = 1\%$ of the evolving incident stellar power and varying heating depth. We find that depending on the heating depth, the radius evolution of hot Jupiters while their stars are on the main-sequence can be classified into three regimes. With shallow heating {that does not extend below pressures of $\sim 10~\mathrm{bars}$,} heating does not greatly affect the radius and the planet perpetually cools --- this is analogous to regime 2(d) of \cite{Komacek:2017a}. With moderately deep heating at $10^2~\mathrm{bars} \lesssim P_\mathrm{dep} \lesssim 10^3~\mathrm{bars}$, heating delays planetary cooling (as in regime 2(c) of \citealp{Komacek:2017a}) but does not cause main-sequence re-inflation. In the case of deep heating at pressures $\gtrsim 10^4~\mathrm{bars}$ (analogous to regimes 2(a) and 2(b) of \citealp{Komacek:2017a}), main-sequence re-inflation can occur. Note that the boundary between the moderately deep heating regime with $10^2~\mathrm{bars} \lesssim P_\mathrm{dep} \lesssim 10^3~\mathrm{bars}$ and the deep heating regime with $P_\mathrm{dep} \gtrsim 10^4~\mathrm{bars}$ depends on the host stellar type -- in principle, cases with $P_\mathrm{dep} \gtrsim 10^2~\mathrm{bars}$ will re-inflate if stellar main-sequence evolution timescales are long enough. In the case of heating at the very center of the planet, main-sequence re-inflation can be significant, with a $\sim 30\%$ increase in the planetary radius over the main-sequence lifetime of the host star for $\gamma = 1\%$.
\begin{figure}
\includegraphics[width=0.5\textwidth]{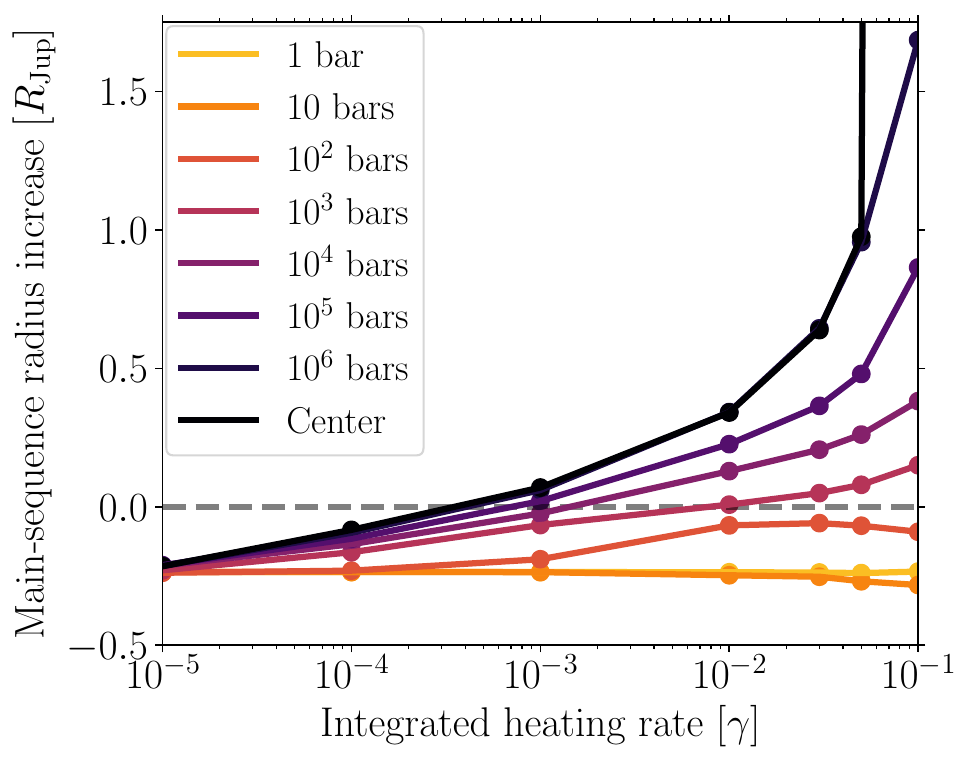}
\caption{\textbf{Re-inflation of hot Jupiters during stellar main-sequence evolution can only occur with heating deeper than 1 kbar.} Shown is the change in planetary radius over the main-sequence stellar evolution phase of the host star from simulations with varying heating rate and deposition pressure. The horizontal dashed line denotes a radius change of zero. Planets below this line shrink over their host stars' main-sequence evolution, while planets above this line re-inflate over main-sequence evolution. We find that heating must be applied at pressures $\gtrsim 10^3~\mathrm{bars}$ to re-inflate planets over the Solar main-sequence. If heating occurs at the very center of the planet, the heating rate must be $\gtrsim 0.1\%$ of the incident stellar power to cause main-sequence re-inflation.}
\label{fig:deltar_ms}
\end{figure}
\begin{figure*}
\begin{center}
\includegraphics[width=1\textwidth]{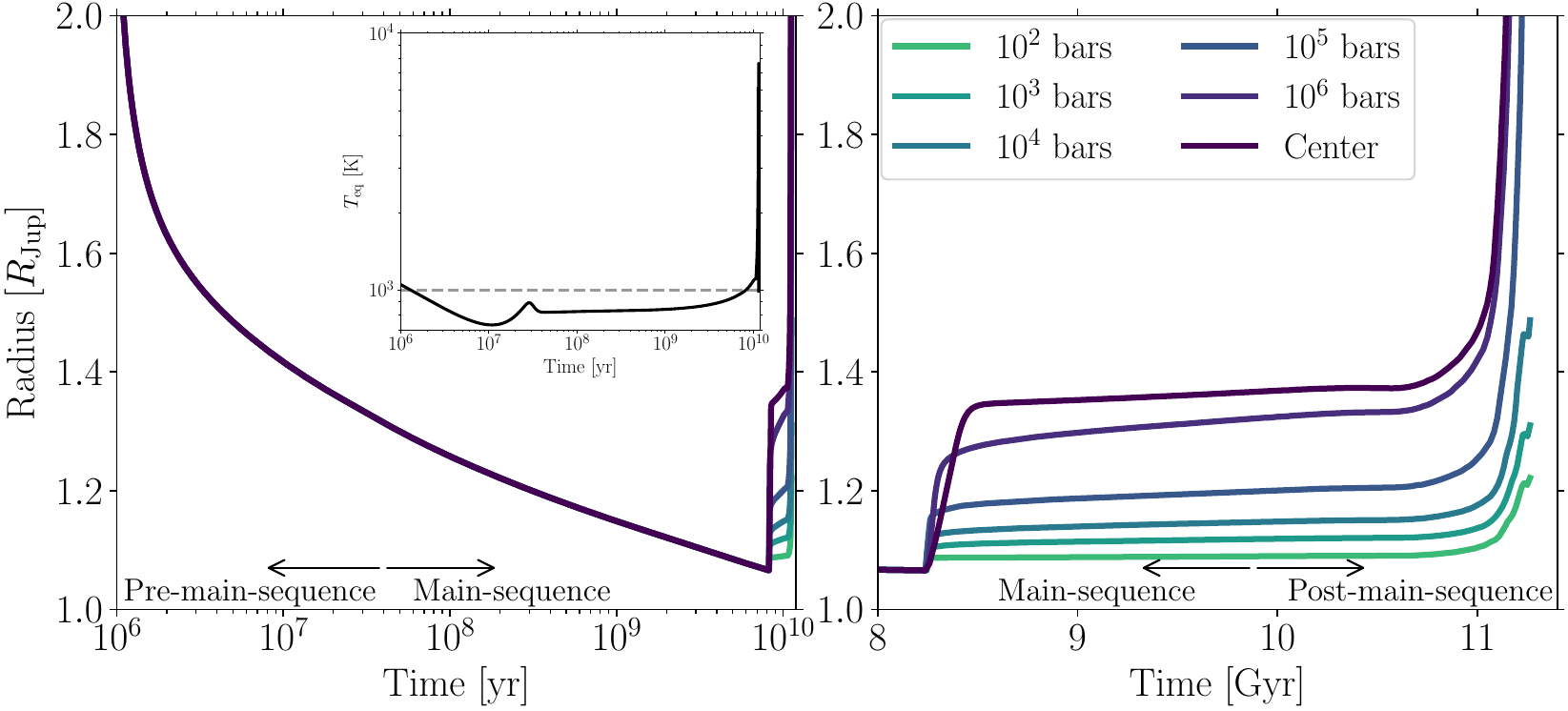}
\caption{\textbf{Post-main-sequence stellar evolution leads to abrupt inflation of warm Jupiters.} The left hand panel shows the radius evolution of a warm Jupiter orbiting at $0.1~\mathrm{au}$ from its host star for varying deposition pressures from $100~\mathrm{bars}$ to the center and a fixed heating rate of $\gamma=1\%$. The inset shows the corresponding equilibrium temperature evolution using MIST Solar evolution tracks \citep{Choi:2016aa,Dotter:2016aa}, and the dashed line in the inset shows the $T_\mathrm{eq} = 1000~\mathrm{K}$ threshold above which deposited heating occurs. The right hand panel shows late evolutionary stages in which planets become re-inflated. The distinction between pre-main-sequence and main-sequence evolutionary stages is shown by the arrows in the left hand panel, and the main-sequence and post-main-sequence phases are marked by arrows in the right hand panel. In this set of models, we assume that heating only occurs when $T_\mathrm{eq} \ge 1000~\mathrm{K}$, as warm Jupiters are observed to not have inflated radii \citep{Miller:2011,Thorngren:2017}. There are two increases in radius after $8$ Gyr: the first is due to the equilibrium temperature reaching $1000~\mathrm{K}$, at which point the heating mechanism turns on, and the second occurs as the star brightens on the post-main-sequence. We confirm the results of \cite{Lopez:2015} that deep heating can significantly re-inflate warm Jupiters. We also find that relatively shallow heating at pressures $\gtrsim 100~\mathrm{bars}$ can lead to significant re-inflation.}
\label{fig:time_pms}
\end{center}
\end{figure*}
\\ \indent The main-sequence radius increase from our full suite of simulations with varying integrated heating rate and heating depth is shown in \Fig{fig:deltar_ms}. We quantify the main-sequence radius increase as the increase in planetary radius between the end of the pre-main-sequence at $39.75~\mathrm{Myr}$ and the end of main-sequence stellar evolution at $9.88~\mathrm{Gyr}$. We find that heating at the center of the planet leads to main-sequence re-inflation if the integrated heating rate $\gamma \gtrsim 0.1\%$. We also find that shallower heating at pressures $P_\mathrm{dep} \ge 10^3~\mathrm{bars}$ can lead to main-sequence re-inflation, given sufficiently strong heating rates of $\gamma \ge 1\%$. The heating rates needed to explain main-sequence re-inflation from our model suite are consistent with the $0.1\% \lesssim \gamma \lesssim 3\%$ heating efficiency needed explain the sample of hot Jupiters with central heat deposition found by \cite{Thorngren:2017}. We will directly incorporate the prescription of \cite{Thorngren:2017} to show that main-sequence re-inflation can  be explained using their derived heating rates in \Sec{sec:test}. 
\subsection{Post-main-sequence re-inflation}
\label{sec:pms}
\begin{figure}
\includegraphics[width=0.49\textwidth]{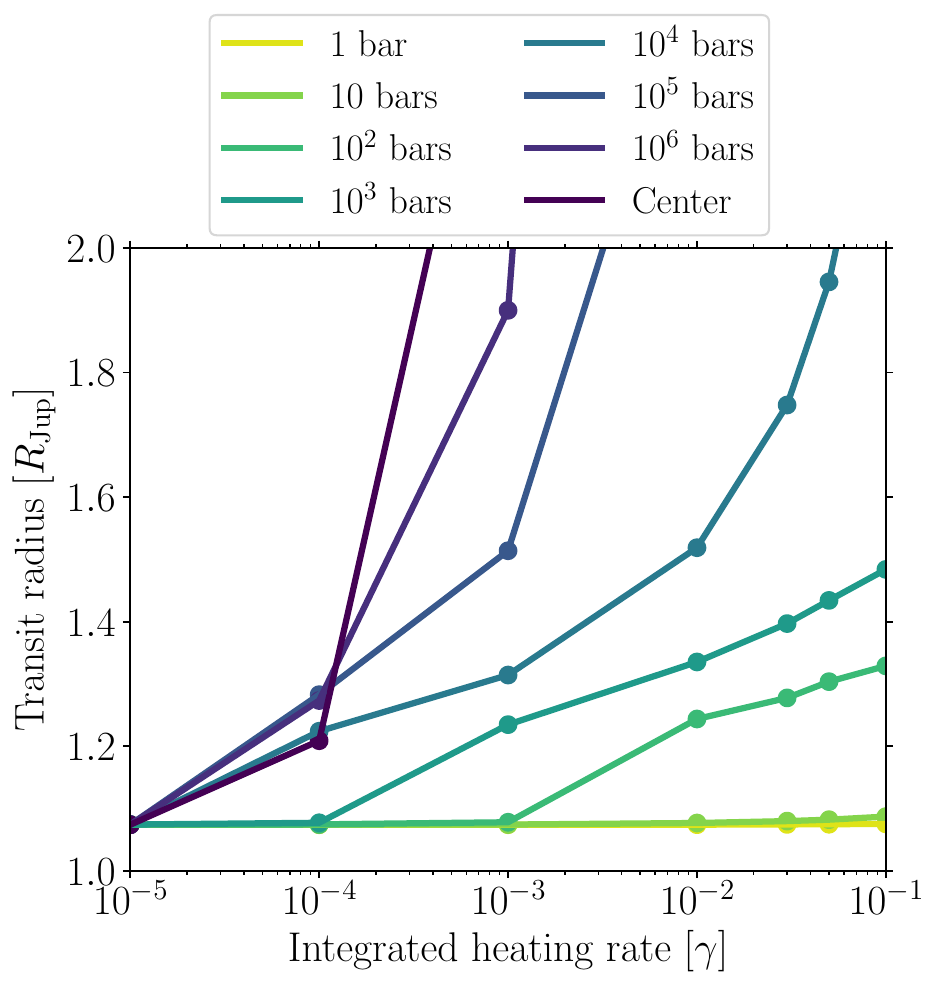}
\caption{\textbf{Warm Jupiters that undergo deep heating will greatly re-inflate during post-main-sequence stellar evolution.} Shown are the transit radii in units of Jupiter radius for warm Jupiters with the mass of HD 209458b orbiting a Sun-like star at a semi-major axis of $0.1~\mathrm{au}$. The transit radii are shown at the time when the host star has evolved to a radius of $10~R_\varodot$. The transit radii are shown for varying integrated heating rates ($\gamma = \Gamma/L_\mathrm{irr}$, from $10^{-3}\%$ to $10\%$) and heating locations ($P_\mathrm{dep}$, from $1~\mathrm{bar}$ to the planet center). We find that heating at pressures $\ge 100~\mathrm{bars}$ is required for re-inflation, while deep heating at pressures $\gtrsim 10^4~\mathrm{bars}$ can lead to a more than doubled radius during post-main-sequence stellar evolution.}
\label{fig:radii_pms}
\end{figure}
\indent Lastly, we show results from our suite of models studying the re-inflation of warm Jupiters while their host stars are on the post-main-sequence. \Fig{fig:time_pms} shows radius evolution tracks for simulations with an integrated heating rate of $\gamma = 1\%$ of the evolving incident stellar power and varying heat deposition pressure. The planet cools over the first $8~\mathrm{Gyr}$ of evolution, after which the equilibrium temperature of the planet is $\ge 1000~\mathrm{K}$ (see inset in the left-hand panel of \Fig{fig:time_pms}) and the heating mechanism turns on. We find that deep heating at  $P_\mathrm{dep} \ge 10^6~\mathrm{bars}$ leads to rapid re-inflation during the late main-sequence (after $T_\mathrm{eq}$ reaches $1000~\mathrm{K}$) with a large increase in the radius during the post-main-sequence evolution of the host star. For moderate heating depths $10^2~\mathrm{bars} \lesssim P_\mathrm{dep} \lesssim 10^5~\mathrm{bars}$, there is only modest re-inflation during the late main-sequence phase where $T_\mathrm{eq} \ge 1000~\mathrm{K}$ but a rapid increase in the planetary radius occurs as the star brightens and approaches the stopping point of $10~R_\varodot$. For shallow heating at pressures $\le 10~\mathrm{bars}$, post-main-sequence re-inflation does not occur (not shown). {Overall, we find that the radius of the planet is tightly linked to the evolving incident stellar flux from the host star, as we found in \Sec{sec:mainseq} for the case of main-sequence re-inflation.} \\
\indent \Fig{fig:radii_pms} shows the transit radius when the host star reaches $10~R_\varodot$ from our full suite of simulations of warm Jupiters with varying integrated heating rate and depth. We find that deep heating can greatly re-inflate planets, even with relatively weak heating rates of $\gamma \lesssim 0.1\%$. With stronger heating, deep heating at $P_\mathrm{dep} \gtrsim 10^5~\mathrm{bars}$ can lead to a runaway in planetary radius, leading to Roche lobe overflow \citep{Valsecchi:2015,Jackson:2017aa}, as found by \cite{Batygin_2011}. {This is also why the cases with heating at pressures $\ge 10^5~\mathrm{bars}$ shown in \Fig{fig:time_pms} inflate to larger than $2~R_\mathrm{Jup}$.} These large radii are caused by the positive feedback between planetary radius, incident stellar power, and deposited heating rate. Larger planets receive more incident stellar power for a given incident stellar flux, which leads to larger deposited heating rates assuming a fixed conversion of incident stellar power to deposited heat. These larger heating rates lead to an increase in the planetary radius, which feeds back and increases the heating rate further, causing a runaway in the planetary radius. {Note that we show in \Sec{sec:test} that this runaway likely would not occur if the deposited heating peaks at an intermediate value of incident flux, as expected for the sample of hot Jupiters \citep{Thorngren:2017}.} \\
\indent \Fig{fig:radii_pms} also shows that relatively shallow heating at pressures $10^2~\mathrm{bars} \lesssim P_\mathrm{dep} \lesssim 10^4~\mathrm{bars}$ with high heating rates $\gamma \gtrsim 1\%$ can lead to the same radius as deep heating at $P_\mathrm{dep} \gtrsim 10^5~\mathrm{bars}$ with weak heating rates $\gamma \lesssim 0.1\%$. At face value, this implies that the degeneracy between the heating rate and heating depth still applies in the case of post-main-sequence re-inflation. However, we will discuss in \Sec{sec:pmsdisc} how this degeneracy can be broken by also considering the evolutionary stage of the host star.
\section{Re-inflation by point source energy deposition}
\label{sec:theory}
\begin{figure*}
\includegraphics[width=1\textwidth]{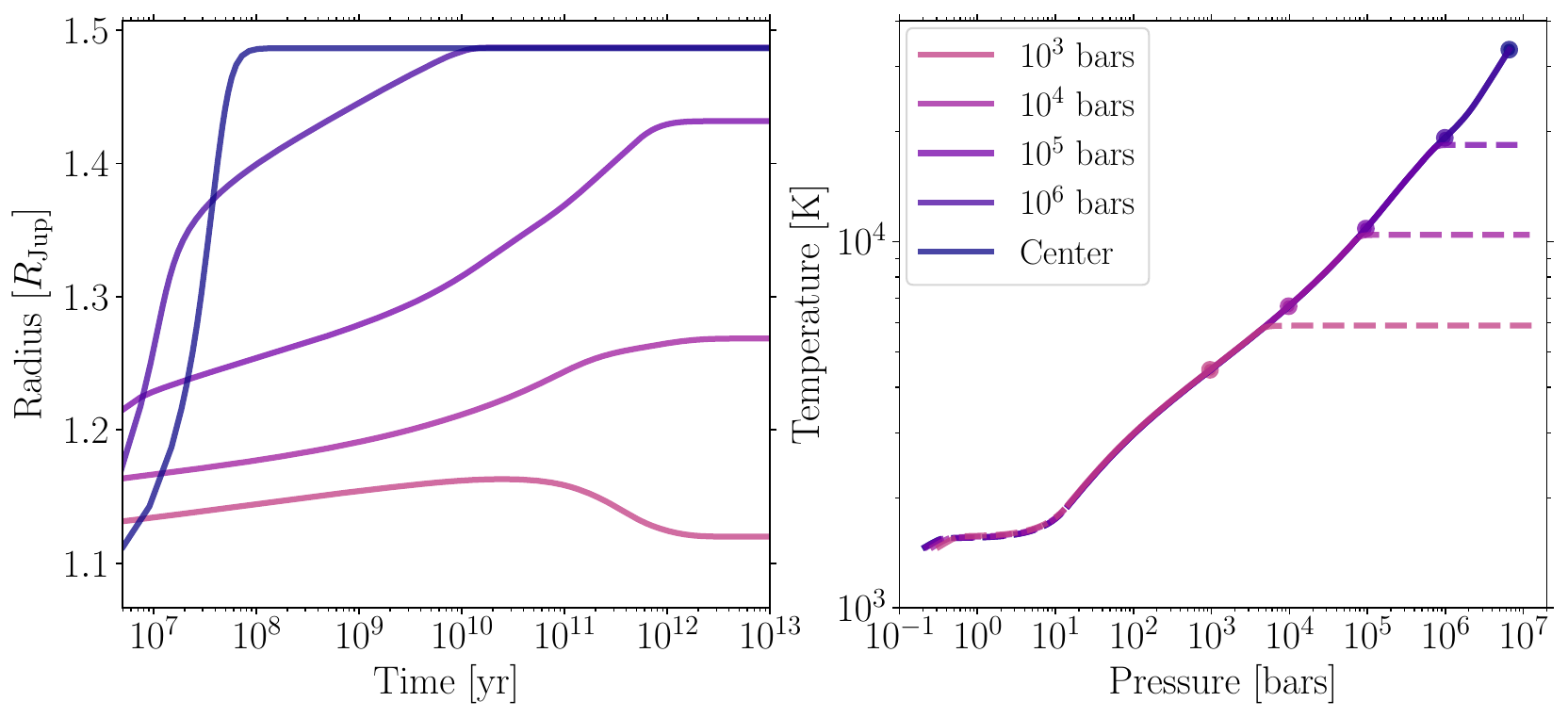}
\caption{\textbf{Re-inflation can take tens to thousands of Gyr if heating is not at the center of the planet, where the final equilibrium includes a deep isotherm from the bottom of the heating level to the center of the planet.} Left hand panel: the radius evolution from simulations with $\gamma = 1\%$ and varying $P_\mathrm{dep}$ from $10^3$ bars to the planet center. This panel shows the continued evolution of simulations from \Fig{fig:rad_age} out to $10~\mathrm{Tyr}$. Simulations with heating at $P_\mathrm{dep} \le 10^5~\mathrm{bars}$ take $\sim 1~\mathrm{Tyr}$ or longer to reach an equilibrium state. 
Right hand panel: the temperature-pressure profiles from the end state of the same simulations as shown in the radius evolution tracks. Solid lines show adiabatic regions, while dashed lines correspond to {non-convective} regions. Points show the maximal heating locations for each $P_\mathrm{dep}$. The deep structure of each case is characterized by an isotherm leading from the bottom of the heating level to the center of the planet. This equilibrium state from re-inflation is the same as in the case of a planet that undergoes delayed cooling due to deposited heating \citep{Ginzburg:2015,Ginzburg:2015a}.}
\label{fig:rad_tp_longruns}
\end{figure*}
\indent To interpret our results, we consider the analytic theory of \cite{Ginzburg:2015,Ginzburg:2015a} for the structure of a planet heated by energy that is deposited at a point within the planetary interior. In this theory, we assume that a heating luminosity $\Gamma$ is deposited at an optical depth $\tau_\mathrm{dep}$. This is a simplification of the actual heating profiles in our numerical simulations, but as we will show accurately reproduces the key features of our numerical results. Additionally, we parameterize the convective profile as in \cite{Ginzburg:2015}:
\begin{equation}
\label{eq:Upower}
\frac{U}{U_c} = \left(\frac{\tau}{\tau_c}\right)^\beta \mathrm{,}
\end{equation}
where $U = a_\mathrm{rad}T^4$ is the radiative energy density with $a_\mathrm{rad}$ the radiation constant. $U_c$ and $\tau_c$ are the radiation energy density and optical depth at the center of the planet, respectively, and $\beta$ is  related to the opacity profile and planetary structure as shown in Equation (3) of \cite{Ginzburg:2015}. \\
\indent We further consider the final end-state at $t = \infty$, which we term the ``equilibrium'' stage of planetary evolution, at which point the planetary structure is in a steady state. This equilibrium state is Stage 4 in the evolution under deposited heating described in Appendix A of \cite{Ginzburg:2015a}. Figure 7 of \cite{Ginzburg:2015a} shows the expected temperature profile at equilibrium. This temperature profile is radiative and nearly isothermal from the outside to the outer radiative-convective boundary located at $\tau_\mathrm{rcb} = 1/\gamma$ (Equation 11 of \citealp{Ginzburg:2015}),
follows the convective power-law profile in \Eq{eq:Upower} from $\tau_\mathrm{rcb}$ to the heating location $\tau_\mathrm{dep}$, and is isothermal from below the heating level to the center of the planet. At equilibrium, the central temperature $T_c$ is set by the heating rate $\gamma$ and depth $\tau_\mathrm{dep}$ and is given by Equation (25) of \cite{Ginzburg:2015}:
\begin{equation}
\label{eq:tc}
\frac{T_c}{T_\mathrm{eq}} \sim \left(1 + \gamma \tau_\mathrm{dep}\right)^{\beta/4} \mathrm{.}
\end{equation}
For inflation that is small compared to the initial size of the planet, the increase in radius $\Delta R$ is directly proportional to the central temperature (see Equation 29 of \citealp{Ginzburg:2015}). As a result, using \Eq{eq:tc} we can derive a scaling for the dependence of the increase in radius on heating rate $\gamma$ and heating depth $\tau_\mathrm{dep}$:
\begin{equation}
\label{eq:theoryprediction}
\Delta R \propto \left(\gamma \tau_\mathrm{dep}\right)^{\beta/4} \mathrm{.}
\end{equation} 
\\ \indent To compare the analytic theory described above to our numerical results, we extend our idealized simulations of re-inflation from \Sec{sec:idealized} out to $10~\mathrm{Tyr}$, at which point the simulations with heating at $P_\mathrm{dep} \ge 10^3~\mathrm{bars}$ reach a final equilibrium. \Fig{fig:rad_tp_longruns} shows the radius evolution and final temperature-pressure profiles of a subset of these simulations with $\gamma = 1\%$ and $P_\mathrm{dep} \ge 10^3~\mathrm{bars}$. We find that all simulations shown reach radius equilibrium by $10~\mathrm{Tyr}$. Cases with $\gamma = 1\%$ and $P_\mathrm{dep} \le 10^2~\mathrm{bars}$ cool below the limits of the \mesa \ equation of state\footnote{This is the region labeled ``here be dragons'' in Figure 50 of \cite{Paxton:2019aa}.}{, as their central temperatures drop below $\sim 5,000~\mathrm{K}$ after Tyrs of evolution. As a result, simulations with shallow heating do not reach equilibrium, and we do not compare them to our analytic theory.}
\\ \indent The temperature-pressure profiles in \Fig{fig:rad_tp_longruns} are characterized by a nearly isothermal outer radiative zone, a convective zone which extends from the radiative-convective boundary to the bottom of the heating level, and an inner radiative zone that is isothermal from the bottom of the heating level to the center of the planet. Planets reach this final structure through re-heating both from the heating level outward toward the surface and from the heating level downward toward the center. \Fig{fig:tp_inside_out} in Appendix \ref{sec:appendixa} shows that the ``inside-out'' heating that leads to the formation of a convective region from the outer radiative-convective boundary to the heating level occurs quickly (within 1 Myr). Inside-out re-inflation is unique to the case of point-source heat deposition, as re-inflation due to heating that decays from the surface inward as considered in \cite{Wu:2013} and \cite{Ginzburg:2015a} only leads to outside-in re-inflation. The equilibrium structure from our numerical simulations is the same structure as was predicted by \cite{Ginzburg:2015a} to occur at the equilibrium stage of planetary evolution. As a result, the final state of hot Jupiters that undergo re-inflation is the same as the final state of hot Jupiters that undergo delayed cooling due to deposited heating. 
\begin{figure}
\includegraphics[width=0.49\textwidth]{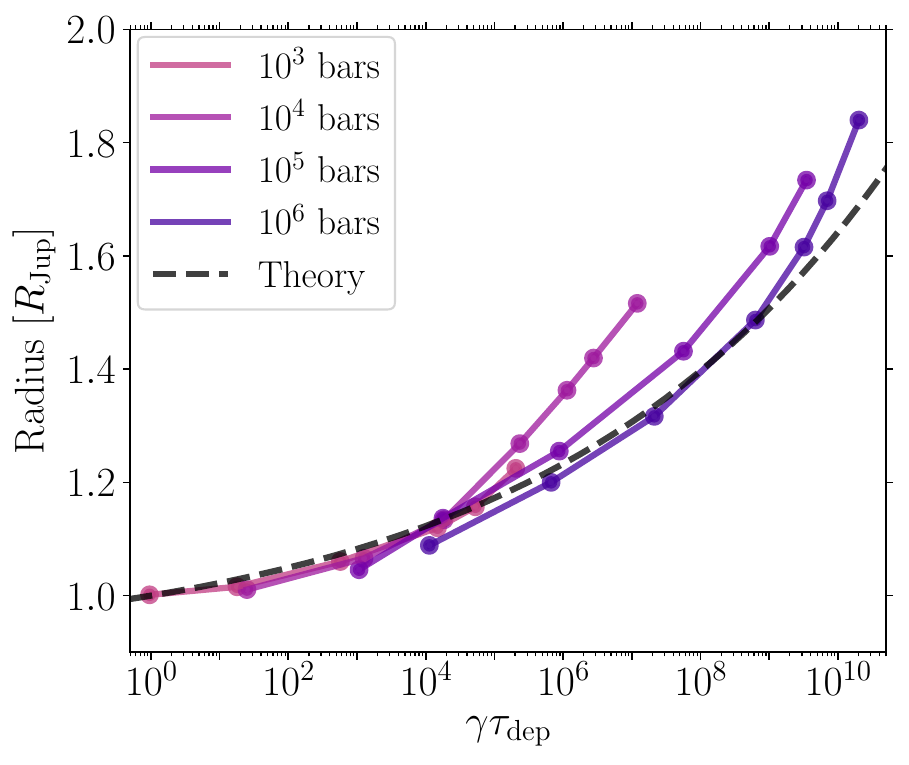}
\caption{\textbf{The analytic theory of \cite{Ginzburg:2015} captures the dependence of equilibrium radius on heating rate and depth found in our suite of idealized models of re-inflation.} Solid lines that connect points show the equilibrium radius from our numerical simulations as a function of the product of the normalized heating rate $\gamma$ and optical depth of the maximal heating location $\tau_\mathrm{dep}$. The dashed line shows our analytic prediction for the dependence of radius on $\gamma \tau_\mathrm{dep}$ from Equation (\ref{eq:theoryprediction}). We find that the analytic prediction agrees with the general trend of increasing radius with increasing $\gamma \tau_\mathrm{dep}$ found in the end-state of our numerical simulations. }
\label{fig:theorycomp}
\end{figure}
\\ \indent We compare our theoretical scaling for the dependence of the equilibrium radius on $\gamma \tau_\mathrm{dep}$ from \Eq{eq:theoryprediction} to that at the final state of our numerical simulations with varying $\gamma$ and $P_\mathrm{dep}$ in \Fig{fig:theorycomp}. We calculate $\beta$ from our numerical simulations, finding that $\beta = 0.348$, in agreement with the value of $0.35$ expected from \cite{Ginzburg:2015}.
{As discussed above,} we do not include simulations with $P_\mathrm{dep} \le 10^2~\mathrm{bars}$ in this comparison {because they do not reach a final equilibrium state in the simulated time frame}. We find that the analytic scaling broadly matches the numerical results for the dependence of radius on the product $\gamma \tau_\mathrm{dep}$. This differs from the results of \cite{Komacek:2017a} (see their Figure 10), where the dependence of radius on $\gamma \tau_\mathrm{dep}$ was not uniform with $P_\mathrm{dep}$. This is because our re-inflation models are evolved to a true equilibrium state, while the comparison with the delayed cooling models of \cite{Komacek:2017a} was done after $5~\mathrm{Gyr}$ of evolution, before the final equilibrium state is reached. As a result, the theory of \cite{Ginzburg:2015,Ginzburg:2015a} can be used to determine the planetary structure for the final equilibrium state at $t = \infty$ {given the combination of heating rate and depth}, {as in the equilibrium state} the heating rate and depth together set the central temperature and radius of the planet. After 10 Gyr of evolution, only some models with deep heating at $P_\mathrm{dep} \gtrsim 10^6~\mathrm{bars}$ reach this equilibrium, while others with shallower heating are still evolving. The long timescales to reach equilibrium for shallow heating that leads to re-inflation are the cause of the differences we found in \Sec{sec:idealized} between heating that delays planetary cooling and heating that leads to re-inflation.
\section{Discussion}
\label{sec:disc}
\subsection{Main-sequence re-inflation}
\label{sec:mainseqdisc}
\indent A key result from this work is that hot Jupiters evolve along with their host stars. For sufficiently deep and strong heating, we expect the radii of hot Jupiters to increase as their host stars brighten. For heating at the very center of the planet, radii can increase by a factor of two over stellar main-sequence evolution. Due to the long timescales of re-inflation, we find that the greatest amount of main-sequence re-inflation occurs between $1-10~\mathrm{Gyr}$ of evolution. As a result, precise stellar ages (using precise stellar parameters derived from asteroseismology and spectral characterization, e.g., \citealp{Grunblatt2016,Grunblatt:2017aa,Grunblatt:2019aa}) are critical for understanding the mechanism that inflates hot Jupiters. \\
\begin{figure}
\includegraphics[width=0.49\textwidth]{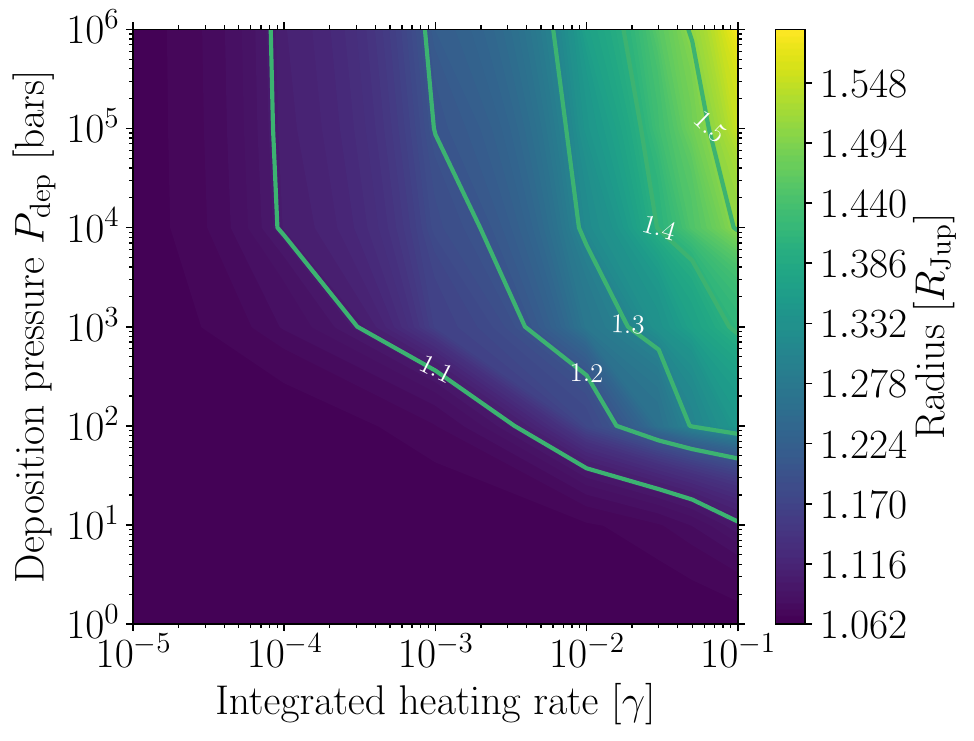}
\caption{\textbf{Warm Jupiters would be significantly inflated if they underwent deep heating during the main-sequence evolution of their host stars.} Contours show the radius at the end of stellar main-sequence evolution for varying heating rate and deposition pressure for a warm Jupiter with the mass of HD 209458b orbiting a Sun-like star at a semi-major axis of $0.1~\mathrm{au}$. In this set of simulations, we do \textit{not} assume that heating only occurs when $T_\mathrm{eq} \ge 1000~\mathrm{K}$, and allow heating to continue below this limit. We find that warm Jupiters are inflated if a heating rate of $\gtrsim 1\%$ of the incident stellar power is deposited deeper than $\sim 10^3~\mathrm{bars}$. The fact that no inflated warm Jupiters have been found \citep{Demory:2011,Miller:2011,Thorngren:2017} means that if warm Jupiters undergo deposited heating, it is too weak and/or too shallow to lead to inflation.}
\label{fig:radii_ms_warm}
\end{figure}
The observation of a lack of inflated warm Jupiters \citep{Demory:2011,Laughlin_2011,Miller:2011,Thorngren:2017,Thorngren:2019aa} points toward weak heating rates and/or shallow heat deposition for planets with $T_\mathrm{eq} < 1000~\mathrm{K}$. {Note that it also might point toward a weaker atmospheric circulation because the planet is not tidally locked, as found by previous studies of the atmospheric circulation of warm Jupiters \citep{Showman:2014,Rauscher:2017aa,Ohno:2019aa}.} To determine the threshold of the combination of heating rate and deposition pressure that would cause warm Jupiters to be inflated, 
we explored the effects of heating on main-sequence evolution of warm Jupiters. To do so, we used the same setup as our main-sequence evolution model suite but studied the evolution of a warm Jupiter at a semi-major axis of $0.1~\mathrm{au}$. Our results for the radius of these warm Jupiters after the main-sequence evolution of a Sun-like star are shown in \Fig{fig:radii_ms_warm}. We find that warm Jupiters would be inflated for integrated heating rates of $\gamma \gtrsim 0.1\%$ and heating depths $P_\mathrm{dep} \gtrsim 10^3~\mathrm{bars}$. \\ 
\indent Because no inflated warm Jupiters have been observed, we infer that the same heating mechanism that inflates hot Jupiters likely does not act to inflate warm Jupiters orbiting main-sequence stars. This finding confirms the validity of our assumption that re-inflated warm Jupiters orbiting post-main-sequence stars are not inflated while their host stars were on the main-sequence.
The weak deposited heating in warm Jupiters agrees with the inferred decrease in deposited heating rate for hot Jupiters at low incident stellar flux \citep{Thorngren:2017}. {This is additional evidence that the radii of close-in gas giant planets are directly tied to the evolution of their host stars through changes in the incident stellar flux.} Additionally, the lack of inflation of warm Jupiters orbiting main-sequence host stars simplifies the interpretation of re-inflated warm Jupiters orbiting post-main-sequence stars, because it is not necessary to determine how inflated the planet was before $T_\mathrm{eq} > 1000~\mathrm{K}$.
\subsection{Post-main-sequence re-inflation}
\label{sec:pmsdisc}
\indent The three candidate re-inflated warm Jupiters orbiting post-main-sequence stars characterized by \cite{Grunblatt2016,Grunblatt:2017aa,Grunblatt:2019aa} all have similar radii of $\approx 1.3-1.45~R_\mathrm{Jup}$ and orbit stars slightly more massive than the Sun. We can explain the radii of these planets in the context of our simulations with either strong heating ($\gamma \sim 1\%$ of the incident stellar flux) that is deposited shallow (at $P_\mathrm{dep} \lesssim 10^4~\mathrm{bars}$) or with weak heating ($\gamma \sim 0.01-0.1\%$) that is deposited deep (at $P_\mathrm{dep} \ge 10^5~\mathrm{bars}$). Our results for the deep heating scenario are consistent with the heating rates required by \cite{Grunblatt:2017aa} to explain the transit radii of K2-97b and K2-132b. \\
\indent Though we find a degeneracy between the inferred heating rate and depth needed to explain re-inflated warm Jupiters, we propose that there are two ways that this degeneracy can be broken. The first is that if heating is deep, we predict that the radii of warm Jupiters will sharply increase as their host star continues to evolve on the post-main-sequence. As a result, if re-inflated warm Jupiters with radii approaching or exceeding $2~R_\mathrm{Jup}$ are detected orbiting evolved post-main-sequence stars, then the heating that causes re-inflation must be deep. The second way to break the degeneracy between heating strength and heating depth is to study the time-evolution of radii of re-inflated warm Jupiters through obtaining precise stellar ages for evolved host stars of re-inflated warm Jupiters. We expect that deep heating is needed to cause rapid re-inflation when the heating mechanism turns on at $T_\mathrm{eq} \ge 1000~\mathrm{K}$. If re-inflated warm Jupiters are found during this late main-sequence phase, then the heating mechanism must be deep. Conversely, if warm Jupiters are not found to be inflated during this late main-sequence phase but are inflated on the post-main-sequence, then the heating must be concentrated at $P_\mathrm{dep} \lesssim 10^5~\mathrm{bars}$. \\
\indent The stellar post-main-sequence evolution timescale decreases for more massive stars. As a result, we expect that heating at different depths will result in different stellar mass distributions for re-inflated warm Jupiters, as less massive stars have longer evolutionary timescales that allow for greater re-inflation. Additionally, there will be a threshold mass above which post-main-sequence re-inflation of warm Jupiters cannot occur due to the short stellar evolution timescales. For central heating, which has the shortest re-inflation timescale of all of our heating depths considered, the heating timescale is $\sim 50~\mathrm{Myr}$ with $\gamma = 1\%$. Complete re-inflation can only occur for warm Jupiters orbiting stars with post-main-sequence lifetimes comparable or longer than the heating timescale. Note that the heating timescale itself will also depend on stellar {class}, because with a fixed conversion of incident stellar power to deposited heating planets orbiting {earlier-type} stars will undergo a larger total heating rate. Additionally, the stellar evolution timescale must be short enough for the host star to reach the post-main-sequence by the present day. Including both these constraints, we expect that re-inflated warm Jupiters will be most prevalent around stars with masses $1 M_\varodot \lesssim M_\star \lesssim 1.5 M_\varodot$. This is the mass range in which current detections of re-inflated warm Jupiters orbiting post-main-sequence stars have been made \citep{Grunblatt2016,Grunblatt:2017aa,Grunblatt:2019aa}. 
\subsection{Using re-inflation to test radius inflation mechanisms}
\label{sec:test}
\begin{figure}
\includegraphics[width=0.49\textwidth]{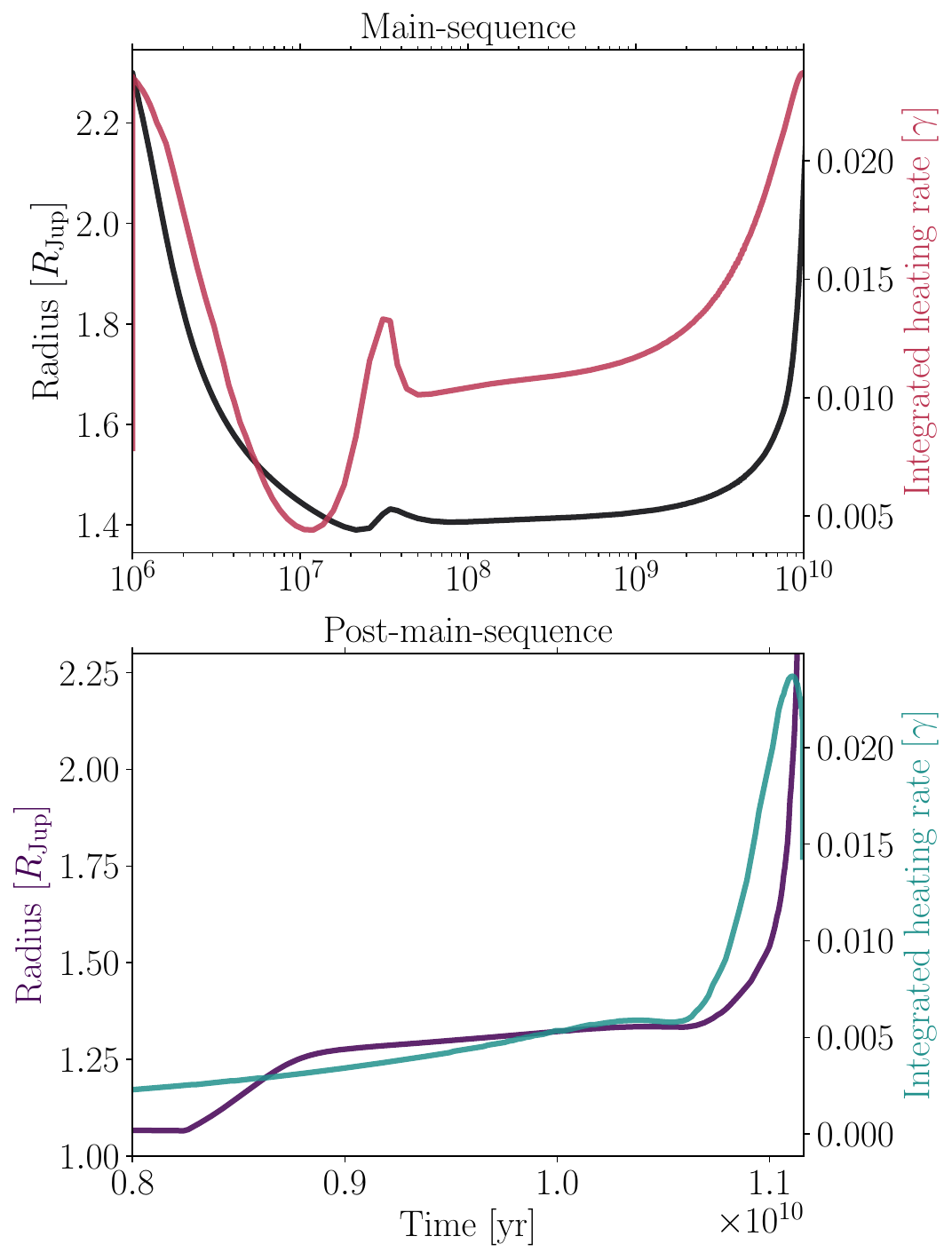}
\caption{\textbf{The heating required to explain the radii of the full sample of hot Jupiters can lead to both main-sequence and post-main-sequence re-inflation.} Shown is the radius evolution for the integrated heating rate inferred by \cite{Thorngren:2017} from the sample of observed hot Jupiters. The radius evolution is shown on the left-hand y-axis, while the heating rate is shown on the right-hand y-axis. The top panel shows main-sequence evolution of hot Jupiters, with a numerical setup similar to our simulations in \Sec{sec:mainseq}. The bottom panel shows the post-main-sequence evolution of warm Jupiters, with a setup similar to that in \Sec{sec:pms}. Note that time is on a logarithmic scale in the top panel and on a linear scale on the bottom panel, which focuses on  post-main-sequence evolution. The heating rate is taken from Equation (34) of \cite{Thorngren:2017} and is Gaussian with a peak at an equilibrium temperature of $\sim 1600~\mathrm{K}$. We find that the dependence of the inferred heating power with incident flux for the full hot Jupiter sample is consistent with both main-sequence re-inflation of hot Jupiters and post-main-sequence re-inflation of warm Jupiters.}
\label{fig:radii_ms_tf18}
\end{figure}
\indent To determine if the inferred heating derived by \cite{Thorngren:2017} from the full sample of hot Jupiters can lead to re-inflation, we ran two additional simulations. One simulation used the same setup as our main-sequence re-inflation suite, while the other used the same setup as our post-main-sequence evolution suite --- the only difference was that in both simulations we used central heating, with the integrated heating rate dependent on the incident stellar flux as in Equation (34) of \cite{Thorngren:2017}. In this model, the heating rate is a Gaussian with a peak at an intermediate value of incident stellar flux that corresponds to an equilibrium temperature of $\approx 1600~\mathrm{K}$. \Fig{fig:radii_ms_tf18} shows the evolution of radius and integrated heating rate from these two simulations. We find that in both simulations the heating rate increases and then decreases as the star brightens. However, the radius remains significantly inflated for both cases.
As a result, we find that the inferred heating rate for the sample of hot Jupiters can explain both main-sequence re-inflation of hot Jupiters and post-main-sequence re-inflation of warm Jupiters. This implies that deep heating mechanisms that weaken in integrated heating rate relative to the incident stellar power at high incident stellar flux may be viable to explain both main-sequence and post-main-sequence re-inflation.   \\
\indent \cite{Thorngren:2019aa} recently showed that the strong heating rates required to explain the radii of hot Jupiters imply that the radiative-convective boundaries of hot Jupiters lie at pressures of $1-100~\mathrm{bars}$, shallower than the $\sim 1~\mathrm{kbar}$ pressures expected from models without additional heating. Such shallow radiative-convective boundaries are consistent with our findings of main-sequence re-inflation, as we expect that inflated planets will have outer radiative-convective boundaries at $\sim 10~\mathrm{bars}$. Additionally, this shallow radiative-convective boundary is consistent with the expectation from simulations of the atmospheric dynamics of hot Jupiters that the deep atmosphere should be nearly adiabatic \citep{Tremblin:2017,Sainsbury-Martinez:2019aa}. \\
\indent In this work, we found that shallow heating at $P_\mathrm{dep} \gtrsim 1~\mathrm{kbar}$ is sufficient to explain main-sequence re-inflation, but that deep heating near the center of the planet is required to explain rapid re-inflation of warm Jupiters. If the heating mechanism leads to deep heating, it can lead to both main-sequence and post-main-sequence re-inflation. However, if the heat is deposited at shallow levels, it will not lead to significant re-inflation of warm Jupiters while the host star is on the main-sequence, even when $T_\mathrm{eq} > 1000~\mathrm{K}$. Additionally, shallow heating will not lead to rapid post-main-sequence re-inflation, and can only lead to inflation up to $\sim 1.5~R_\mathrm{Jup}$ (see Figure \ref{fig:radii_pms}). It is possible that main-sequence and post-main-sequence re-inflation are caused by different heating mechanisms. In this case, the mechanism that causes post-main-sequence re-inflation would lead to deep heat deposition, while the (separate) mechanism that causes main-sequence re-inflation would lead to relatively shallow heat deposition. \\
\indent We can relate the possibility of different heating depths for main-sequence and post-main-sequence re-inflation discussed above  to distinct proposed heating mechanisms. For instance, post-main-sequence re-inflation of warm Jupiters could be due to a non-zero initial eccentricity that enables strong tidal dissipation as the host star evolves off the main-sequence, while main-sequence re-inflation of hot Jupiters could be caused by mechanisms related to the atmospheric circulation (e.g., Ohmic dissipation or an atmospheric heat flux directed inward). This is consistent with the expectation from previous work \citep{Wu:2013,Lopez:2015,Ginzburg:2015a} that Ohmic dissipation will not lead to rapid re-inflation. Additionally, both shallow and deep heating mechanisms could act together to cause re-inflation. Notably, if tidal dissipation provides a deep heat source for warm Jupiters orbiting post-main-sequence stars, we would expect it to occur for only the fraction of planets that still have a non-zero eccentricity as the host star evolves off the main-sequence. This is because tidal damping timescales for warm Jupiters orbiting Sun-like stars are on the order of Gyr \citep{Gu:2003aa,Grunblatt:2017aa}. As a result, we expect that tidal dissipation will not be a ubiquitous process for warm Jupiters orbiting post-main-sequence stars. \\
\indent Future observations of a wide sample of re-inflated warm Jupiters will test mechanisms for radius inflation. TESS will observe $\sim 400,000$ evolved stars, with an expected $0.51 \pm 0.29\%$ occurrence rate of close-in re-inflated warm Jupiters around post-main-sequence stars \citep{Grunblatt:2019aa}. As a result, we expect that TESS will discover a large sample of
of re-inflated warm Jupiters. This large sample will directly test how deep deposited heating needs to be to re-inflate warm Jupiters. If heating occurs near the center of the planet, warm Jupiters will undergo fast re-inflation and TESS will find highly inflated planets with radii approaching the Roche limit. If heating is instead relatively shallow, there will be a lack of highly inflated planets and TESS will find that the occurrence rate of re-inflated planets increases sharply as the radii of host stars approach $10~R_\varodot$. 

\section{Conclusions}
\label{sec:conc}
In this work, we studied how deposited heating leads to re-inflation of hot Jupiters. To do so, we used \mesa \ to compute three suites of planetary evolution models: one to elucidate the process by which planets re-inflate, a second studying hot Jupiter evolution with deposited heating over the main-sequence evolution of their host star, and third studying the post-main-sequence re-inflation of warm Jupiters. We found that deposited heating can lead to both main-sequence re-inflation of hot Jupiters and post-main-sequence re-inflation of warm Jupiters, provided it is deep enough and has a sufficient dissipation rate. Our key conclusions are as follows:
\begin{enumerate}
\item Deeper heating is required to re-inflate planets to a given radius after billions of years of evolution than for the planet to reach the same radius through heating that delays planetary cooling. This is because re-inflation must very slowly heat the interior of the planet from the heating level downward, and does not greatly affect the central temperature unless the heating is deep. As a result, the radius of a planet after re-inflation increases with increasing heating depth and increasing heating rate, with central heating required to lead to maximum re-inflation. We compared the analytic theory of \cite{Ginzburg:2015,Ginzburg:2015a} for the equilibrium radius and temperature profile of planets that have undergone re-inflation to our numerical simulations, finding good agreement throughout the range of heating rates and deposition pressures considered.
\item There is a strong degeneracy between the deposited heating rate and depth that complicates the interpretation of hot Jupiters that are re-inflated during the main-sequence evolution of their host stars. As a result, a range of heating profiles can explain main-sequence re-inflation of hot Jupiters, including weak heating of $\approx 0.1\%$ of the incident stellar flux deposited the very center of the planet and high heating rates of $\gtrsim 1\%$ of the incident stellar flux deposited at a pressure of $\sim 10^3~\mathrm{bars}$.
\item The degeneracy between deposited heating rate and depth can be broken in the case of re-inflated warm Jupiters orbiting post-main-sequence stars. The radii of recently discovered re-inflated warm Jupiters orbiting post-main-sequence stars \citep{Grunblatt2016,Grunblatt:2017aa,Grunblatt:2019aa} can be explained with either weak heating at the center of the planet \citep{Lopez:2015} or strong shallow heating. However, post-main-sequence re-inflation occurs much more rapidly for deep heating, and shallow heating cannot explain re-inflation over late stages of main-sequence host stellar evolution. The large sample of observed re-inflated warm Jupiters orbiting post-main-sequence stars that will be obtained by TESS, combined with precise stellar ages, can determine the depth of the heating source that leads to inflation. 
\item The dependence of the heating rate on incident stellar flux inferred from the sample of hot Jupiters by \cite{Thorngren:2017} can explain both main-sequence re-inflation of hot Jupiters and post-main-sequence re-inflation of warm Jupiters, if heat is deposited at the center of the planet. As a result, the heating rate does not need to have a monotonic dependence on incident stellar flux to lead to re-inflation. We find that heating must be weak for warm Jupiters with equilibrium temperatures $\lesssim 1000~\mathrm{K}$, as otherwise they would be inflated while their host stars are on the main-sequence. The lack of deposited heat in warm Jupiters with $T_\mathrm{eq} < 1000~\mathrm{K}$ orbiting main-sequence stars also agrees with the inferred dependence of the deposited heating rate on incident stellar flux from the hot Jupiter sample. Mechanisms that cause deep heating and decrease in efficacy at low and high incident stellar flux can therefore potentially explain both re-inflation of hot Jupiters orbiting main-sequence stars and re-inflation of warm Jupiters orbiting post-main-sequence stars. 
\end{enumerate}
\begin{figure*}[ht]
\includegraphics[width=1\textwidth]{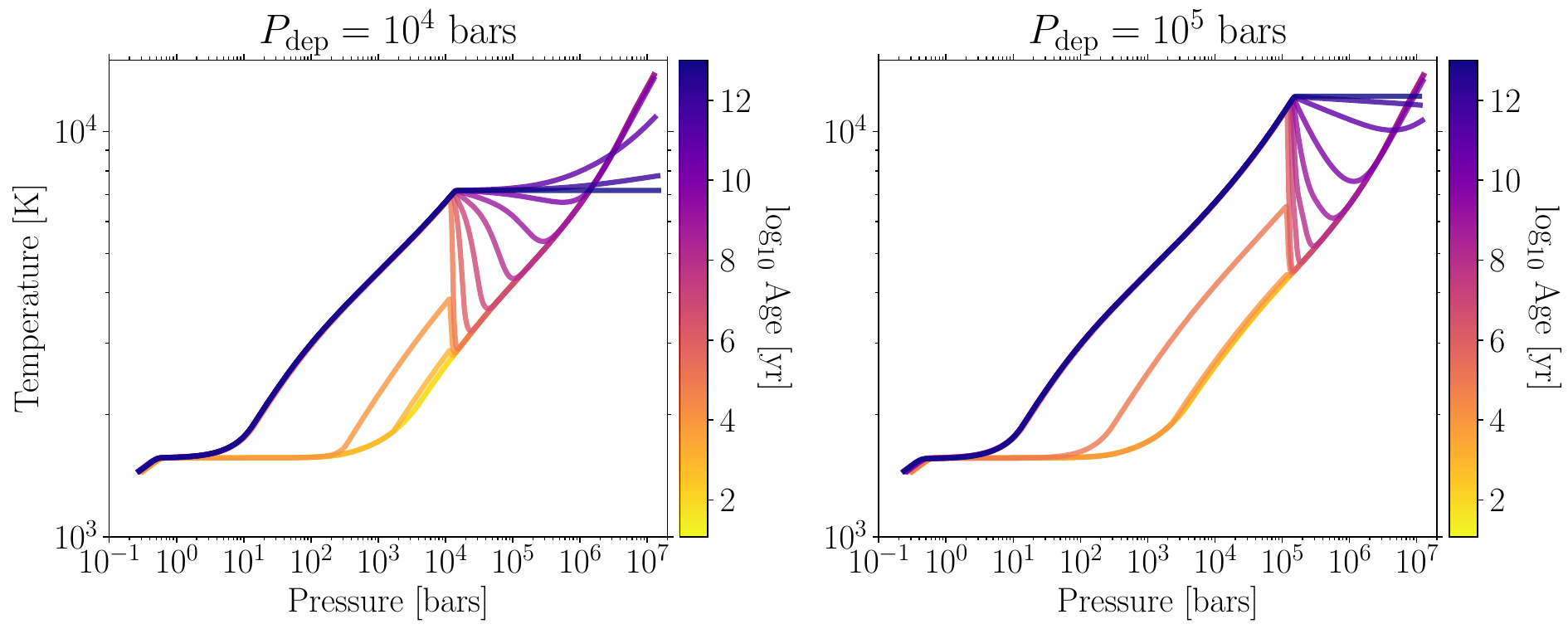}
\caption{\textbf{Point source heat deposition leads to both inside-out and outside-in re-inflation.} Shown are temperature-pressure profiles at varying ages from $10-10^{13}$ yr for simulations similar to those described in \Sec{sec:theory}, with a heating rate of $\gamma = 1\%$ and $P_\mathrm{dep} = 10^4~\mathrm{bars}$ (left panel) and $P_\mathrm{dep} = 10^5~\mathrm{bars}$ (right panel). However, these experiments use a narrower standard deviation of the heating rate of $0.1$ pressure scale heights, rather than the $0.5$ used in our standard cases. We use this narrow heating distribution to more accurately reproduce point-source heating, and to more cleanly show the effect of heating on inside-out and outside-in re-inflation. For visual clarity, we do not identify whether regions are radiative or convective. We find that re-inflation from the heating level upward is rapid, occurring in less than a Myr for both cases. As a result, inside-out re-inflation occurs much more quickly than outside-in re-inflation.}
\label{fig:tp_inside_out}
\end{figure*}
\acknowledgements
We thank Jonathan Fortney for insightful comments on an early draft of this manuscript, and we thank Konstantin Batygin and Andrew Youdin for helpful discussions. We thank the referee for thoughtful comments that improved this work. We also thank the \mesa~team for making this valuable tool publicly available. T.D.K. and S.G. acknowledge support from the 51 Pegasi b Fellowship in Planetary Astronomy sponsored by the Heising-Simons Foundation. D.P.T. acknowledges support by the Trottier Fellowship from the Exoplanet Research Institute (iREx). E.D.L. would like to acknowledge support from the GSFC Sellers Exoplanet Environments Collaboration (SEEC), which is funded in part by the NASA Planetary Science Division’s Internal Scientist Funding Model. This work initiated from discussions at the 2019 Exoplanet Summer Program in the Other Worlds Laboratory (OWL) at the University of California, Santa Cruz, a program funded by the Heising-Simons Foundation. T.D.K. would furthermore like to acknowledge that the bulk of this work was performed on land that is the traditional homeland of the people of the Council of Three Fires, the Ojibwe, Potawatomi, and Odawa as well as the Menominee, Miami, Ho-Chunk, Sac, and Fox nations.
\appendix

\section{Inside-out vs. outside-in re-inflation}
\label{sec:appendixa}
\indent To more clearly display the evolution of a planet undergoing re-inflation from point source heat deposition, we consider a narrower heating profile of a Gaussian with a standard deviation of $0.1$ pressure scale heights, rather than the $0.5$ pressure scale heights used in our nominal grids of simulations. We conducted numerical experiments with this narrowed heating profile for a heating rate of $\gamma = 1\%$ and moderate $P_\mathrm{dep} = 10^4~\mathrm{bars}$ and $10^5~\mathrm{bars}$, and carried them out to $10$ Tyr as in our suite of simulations described in \Sec{sec:theory}. \\\indent \Fig{fig:tp_inside_out} shows the evolution of temperature-pressure profiles from 10 yr to $10^{13}$ yr from these two experiments. We find that in both cases, the outer envelope re-inflates from the heating level outward. The radiative-convective boundary is deep (at $\sim 1~\mathrm{kbar}$) at early times, and evolves outward as the planet re-inflates. This inside-out heating leads to regions above the heating level becoming convective, and reaching a fixed temperature with time by $1$ Myr in both cases. Meanwhile, the interior warms up due to deposited heating over much longer timescales, only reaching a fixed isothermal temperature profile below the heating level by $10$ Tyr. \\
\indent The evolution of our cases with point-like heat deposition at early times differs with expectations from the Ohmic dissipation models of \cite{Wu:2013} and \cite{Ginzburg:2015a}. In the case of Ohmic dissipation {alone}, re-inflation is purely from the heating level downward (i.e., outside-in) because the heating rate decays with increasing pressure and because outer regions of the planet have a lower heat capacity than inner regions. {However, note that vertical motions could transport deposited heat upward, acting as inside-out heating}. For point source heat deposition, heating acts to re-inflate the planet both from the heating level upward (i.e., inside-out) and from the outside-in. However, the timescale of the inside-out heating is rapid ($\lesssim 1$ Myr) relative to the time it takes the planet to re-inflate from the outside-in, which can be $\gtrsim 1$ Tyr for intermediate deposition depths. As a result, the majority of the radius evolution of re-inflated planets undergoing point source heat deposition is determined by the rate of outside-in re-inflation.  

\if\bibinc n
\bibliography{References_all}

\begin{thebibliography}{78}
\expandafter\ifx\csname natexlab\endcsname\relax\def\natexlab#1{#1}\fi

\bibitem[{Arras \& Bildsten(2006)}]{Arras:2006kl}
Arras, P. \& Bildsten, L. 2006, The Astrophysical Journal, 650, 394

\bibitem[{Arras \& Socrates(2010)}]{Arras:2010}
Arras, P. \& Socrates, A. 2010, The Astrophysical Journal, 714, 1

\bibitem[{Baraffe {et~al.}(2010)Baraffe, Chabrier, \& Barman}]{Baraffe:2010xe}
Baraffe, I., Chabrier, G., \& Barman, T. 2010, Reports on Progress in Physics,
  76, 30

\bibitem[{Baraffe {et~al.}(2014)Baraffe, Chabrier, Fortney, \&
  Sotin}]{Baraffe:2014}
Baraffe, I., Chabrier, G., Fortney, J., \& Sotin, C. Protostars and Planets VI,
  ed. H.~Beuther, R.~Klessen, C.~Dullemond, \& T.~Henning (Tucson, AZ:
  University of Arizona Press)

\bibitem[{Batygin \& Stevenson(2010)}]{Batygin_2010}
Batygin, K. \& Stevenson, D. 2010, The Astrophysical Journal Letters, 714, L238

\bibitem[{Batygin {et~al.}(2011)Batygin, Stevenson, \&
  Bodenheimer}]{Batygin_2011}
Batygin, K., Stevenson, D., \& Bodenheimer, P. 2011, The Astrophysical Journal,
  738, 1

\bibitem[{Bodenheimer {et~al.}(2001)Bodenheimer, Lin, \&
  Mardling}]{Bodenheimer:2001}
Bodenheimer, P., Lin, D., \& Mardling, R. 2001, The Astrophysical Journal, 548,
  466

\bibitem[{Budaj {et~al.}(2012)Budaj, Hubeny, \& Burrows}]{Budaj:2012aa}
Budaj, J., Hubeny, I., \& Burrows, A. 2012, Astronomy {\&} Astrophysics, 537,
  A115

\bibitem[{Burrows {et~al.}(2007)Burrows, Hubeny, Budaj, \&
  Hubbard}]{Burrows:2007bs}
Burrows, A., Hubeny, I., Budaj, J., \& Hubbard, W. 2007, The Astrophysical
  Journal, 661, 502

\bibitem[{Chabrier \& Baraffe(2007)}]{Chabrier:2007}
Chabrier, G. \& Baraffe, I. 2007, The Astrophysical Journal Letters, 661, 81

\bibitem[{Chandrasekhar(1939)}]{Chandrasekhar:1939}
Chandrasekhar, S. 1939, An Introduction to Stellar Structure (Chicago, IL:
  University of Chicago Press)

\bibitem[{Choi {et~al.}(2016)Choi, Dotter, Conroy, Cantiello, Paxton, \&
  Johnson}]{Choi:2016aa}
Choi, J., Dotter, A., Conroy, C., Cantiello, M., Paxton, B., \& Johnson, B.
  2016, The Astrophysical Journal, 823, 102

\bibitem[{Demory \& Seager(2011)}]{Demory:2011}
Demory, B. \& Seager, S. 2011, The Astrophysical Journal Supplement Series,
  197, 12

\bibitem[{Dotter(2016)}]{Dotter:2016aa}
Dotter, A. 2016, The Astrophysical Journal Supplement Series, 222, 8

\bibitem[{Fortney {et~al.}(2010)Fortney, Baraffe, \& Militzer}]{fortney_2009}
Fortney, J., Baraffe, I., \& Militzer, B. Exoplanets, ed. S.~Seager (Tucson,
  AZ: University of Arizona Press)

\bibitem[{Fortney {et~al.}(2007)Fortney, Marley, \& Barnes}]{Fortney:2007ta}
Fortney, J., Marley, M., \& Barnes, J. 2007, The Astrophysical Journal, 659,
  1661

\bibitem[{{Fortney} {et~al.}(2008){Fortney}, {Lodders}, {Marley}, \&
  {Freedman}}]{fortney-etal-2008}
{Fortney}, J.~J., {Lodders}, K., {Marley}, M.~S., \& {Freedman}, R.~S. 2008,
  \apj, 678, 1419

\bibitem[{Freedman {et~al.}(2008)Freedman, Marley, \& Lodders}]{Freedman:2008}
Freedman, R., Marley, M., \& Lodders, K. 2008, The Astrophysical Journal
  Supplement Series, 174, 504

\bibitem[{Ginzburg \& Sari(2015)}]{Ginzburg:2015}
Ginzburg, S. \& Sari, R. 2015, The Astrophysical Journal, 803, 111

\bibitem[{Ginzburg \& Sari(2016)}]{Ginzburg:2015a}
---. 2016, The Astrophysical Journal, 819, 116

\bibitem[{Grunblatt {et~al.}(2019)Grunblatt, Huber, Gaidos, Hon, Zinn, \&
  Stello}]{Grunblatt:2019aa}
Grunblatt, S., Huber, D., Gaidos, E., Hon, M., Zinn, J., \& Stello, D. 2019,
  The Astronomical Journal, 158, 227

\bibitem[{Grunblatt {et~al.}(2017)Grunblatt, Huber, Gaidos, Lopez, Howard,
  Isaacson, Sinukoff, Vanderburg, Nofi, Yu, North, Chaplin, Foreman-Mackey,
  Petigura, Ansdell, Weiss, Fulton, \& Lin}]{Grunblatt:2017aa}
Grunblatt, S., Huber, D., Gaidos, E., Lopez, E., Howard, A., Isaacson, H.,
  Sinukoff, E., Vanderburg, A., Nofi, L., Yu, J., North, T., Chaplin, W.,
  Foreman-Mackey, D., Petigura, E., Ansdell, M., Weiss, L., Fulton, B., \& Lin,
  D. 2017, The Astronomical Journal, 154, 254

\bibitem[{Grunblatt {et~al.}(2016)Grunblatt, Huber, Gaidos, Lopez, Fulton,
  Vanderburg, Barclay, Fortney, Howard, \& Isaacson}]{Grunblatt2016}
Grunblatt, S.~K., Huber, D., Gaidos, E.~J., Lopez, E.~D., Fulton, B.~J.,
  Vanderburg, A., Barclay, T., Fortney, J.~J., Howard, A.~W., \& Isaacson,
  H.~T. 2016, The Astronomical Journal, 152, 1

\bibitem[{Gu {et~al.}(2004)Gu, Bodenheimer, \& Lin}]{Gu:2004aa}
Gu, P., Bodenheimer, P., \& Lin, D. 2004, The Astrophysical Journal, 608, 1076

\bibitem[{Gu {et~al.}(2003)Gu, Lin, \& Bodenheimer}]{Gu:2003aa}
Gu, P., Lin, D., \& Bodenheimer, P. 2003, The Astrophysical Journal, 588, 509

\bibitem[{Gu {et~al.}(2019)Gu, Peng, \& Yen}]{Gu:2019aa}
Gu, P., Peng, D., \& Yen, C. 2019, The Astrophysical Journal, 887, 228

\bibitem[{Guillot(2010)}]{Guillot:2010}
Guillot, T. 2010, Astronomy and Astrophysics, 520, A27

\bibitem[{Guillot \& Showman(2002)}]{Guillot_2002}
Guillot, T. \& Showman, A. 2002, Astronomy and Astrophysics, 385, 156

\bibitem[{Hartman {et~al.}(2016)Hartman, Bakos, Bhatti, Penev, Bieryla, Latham,
  Kov{\'{a}}cs, Torres, Csubry, de~Val-Borro, Buchhave, Kov{\'{a}}cs, Quinn,
  Howard, Isaacson, Fulton, Everett, Esquerdo, B{\'{e}}ky, Szklenar, Falco,
  Santerne, Boisse, H{\'{e}}brard, Burrows, L{\'{a}}z{\'{a}}r, Papp, \&
  S{\'{a}}ri}]{Hartman2016}
Hartman, J.~D., Bakos, G.~{\'{A}}., Bhatti, W., Penev, K., Bieryla, A., Latham,
  D.~W., Kov{\'{a}}cs, G., Torres, G., Csubry, Z., de~Val-Borro, M., Buchhave,
  L., Kov{\'{a}}cs, T., Quinn, S., Howard, A.~W., Isaacson, H., Fulton, B.~J.,
  Everett, M.~E., Esquerdo, G., B{\'{e}}ky, B., Szklenar, T., Falco, E.,
  Santerne, A., Boisse, I., H{\'{e}}brard, G., Burrows, A., L{\'{a}}z{\'{a}}r,
  J., Papp, I., \& S{\'{a}}ri, P. 2016, The Astronomical Journal, 152, 182

\bibitem[{Huang \& Cumming(2012)}]{Huang_2012}
Huang, X. \& Cumming, A. 2012, The Astrophysical Journal, 757, 47

\bibitem[{Ibgui \& Burrows(2009)}]{Ibgui:2009}
Ibgui, L. \& Burrows, A. 2009, The Astrophysical Journal, 700, 1921

\bibitem[{Ibgui {et~al.}(2010)Ibgui, Burrows, \& Spiegel}]{Ibgui:2010}
Ibgui, L., Burrows, A., \& Spiegel, D. 2010, The Astrophysical Journal, 713,
  751

\bibitem[{Jackson {et~al.}(2017)Jackson, Arras, Penev, Peacock, \&
  Marchant}]{Jackson:2017aa}
Jackson, B., Arras, P., Penev, K., Peacock, S., \& Marchant, P. 2017, The
  Astrophysical Journal, 835, 145

\bibitem[{Jackson {et~al.}(2008)Jackson, Greenberg, \& Barnes}]{Jackson:681}
Jackson, B., Greenberg, R., \& Barnes, R. 2008, The Astrophysical Journal, 681,
  1631

\bibitem[{Kippenhahn {et~al.}(2012)Kippenhahn, Weigert, \&
  Weiss}]{Kippenhahn:2012}
Kippenhahn, R., Weigert, A., \& Weiss, A. 2012, Stellar Structure and
  Evolution, 2nd edn. (New York: Springer)

\bibitem[{Komacek \& Showman(2016)}]{Komacek:2015}
Komacek, T. \& Showman, A. 2016, The Astrophysical Journal, 821, 16

\bibitem[{Komacek {et~al.}(2019)Komacek, Showman, \&
  Parmentier}]{Komacek:2019aa}
Komacek, T., Showman, A., \& Parmentier, V. 2019, The Astrophysical Journal,
  881, 152

\bibitem[{Komacek {et~al.}(2017)Komacek, Showman, \& Tan}]{Komacek:2017}
Komacek, T., Showman, A., \& Tan, X. 2017, The Astrophysical Journal, 835, 198

\bibitem[{Komacek \& Youdin(2017)}]{Komacek:2017a}
Komacek, T. \& Youdin, A. 2017, The Astrophysical Journal, 844, 94

\bibitem[{Kurokawa \& Inutsuka(2015)}]{Kurokawa:2015aa}
Kurokawa, H. \& Inutsuka, S. 2015, The Astrophysical Journal, 815, 78

\bibitem[{Laughlin(2018)}]{Laughlin:2018aa}
Laughlin, G. The Exoplanet Handbook (Springer)

\bibitem[{Laughlin {et~al.}(2011)Laughlin, Crismani, \& Adams}]{Laughlin_2011}
Laughlin, G., Crismani, M., \& Adams, F. 2011, The Astrophysical Journal
  Letters, 729, L7

\bibitem[{Laughlin \& Lissauer(2015)}]{Laughlin:2015}
Laughlin, G. \& Lissauer, J. Treatise on Geopysics, 2nd edn., ed. G.~Schubert
  (Elsevier)

\bibitem[{Leconte \& Chabrier(2012)}]{Leconte:2012}
Leconte, J. \& Chabrier, G. 2012, Astronomy and Astrophysics, 540, A20

\bibitem[{Leconte {et~al.}(2010)Leconte, Chabrier, Baraffe, \&
  Levrard}]{Leconte:2010a}
Leconte, J., Chabrier, G., Baraffe, I., \& Levrard, B. 2010, Astronomy and
  Astrophysics, 516, A64

\bibitem[{Lopez \& Fortney(2016)}]{Lopez:2015}
Lopez, E. \& Fortney, J. 2016, The Astrophysical Journal, 818, 4

\bibitem[{Menou(2012)}]{Menou:2012fu}
Menou, K. 2012, The Astrophysical Journal, 745, 138

\bibitem[{Miller \& Fortney(2011)}]{Miller:2011}
Miller, N. \& Fortney, J. 2011, The Astrophysical Journal Letters, 736, L29

\bibitem[{Miller {et~al.}(2009)Miller, Fortney, \& Jackson}]{Miller:2009}
Miller, N., Fortney, J., \& Jackson, B. 2009, The Astrophysical Journal, 702,
  1413

\bibitem[{Millholland(2019)}]{Millholland:2019aa}
Millholland, S. 2019, The Astrophysical Journal, 886, 72

\bibitem[{Ohno \& Zhang(2019)}]{Ohno:2019aa}
Ohno, K. \& Zhang, X. 2019, The Astrophysical Journal, 874, 1

\bibitem[{Owen \& Wu(2016)}]{Owen:2015}
Owen, J. \& Wu, Y. 2016, The Astrophysical Journal, 817, 107

\bibitem[{Paxton {et~al.}(2011)Paxton, Bildsten, Dotter, Herwig, Lesaffre, \&
  Timmes}]{Paxton:2011}
Paxton, B., Bildsten, L., Dotter, A., Herwig, F., Lesaffre, P., \& Timmes, F.
  2011, The Astrophysical Journal Supplement Series, 192, 3

\bibitem[{Paxton {et~al.}(2013)Paxton, Cantiello, Arras, Bildsten, Brown,
  Dotter, Mankovich, Montgomery, Stello, Timmes, \& Townsend}]{Paxton:2013}
Paxton, B., Cantiello, M., Arras, P., Bildsten, L., Brown, E., Dotter, A.,
  Mankovich, C., Montgomery, M., Stello, D., Timmes, F., \& Townsend, R. 2013,
  The Astrophysical Journal Supplement Series, 208, 4

\bibitem[{Paxton {et~al.}(2015)Paxton, Marchant, Schwab, Bauer, Bildsten,
  Cantiello, Dessart, Farmer, Hu, Langer, Townsend, Townsley, \&
  Timmes}]{Paxton:2015}
Paxton, B., Marchant, P., Schwab, J., Bauer, E., Bildsten, L., Cantiello, M.,
  Dessart, L., Farmer, R., Hu, H., Langer, N., Townsend, R., Townsley, D., \&
  Timmes, F. 2015, The Astrophysical Journal Supplement Series, 220, 15

\bibitem[{Paxton {et~al.}(2018)Paxton, Schwab, Bauer, Bildsten, Blinnikov,
  Duffell, Farmer, Goldberg, Marchant, Sorokina, Thoul, Townsend, \&
  Timmes}]{Paxton:2018aa}
Paxton, B., Schwab, J., Bauer, E., Bildsten, L., Blinnikov, S., Duffell, P.,
  Farmer, R., Goldberg, J., Marchant, P., Sorokina, E., Thoul, A., Townsend,
  R., \& Timmes, F. 2018, The Astrophysical Journal Supplement Series, 234, 34

\bibitem[{Paxton {et~al.}(2019)Paxton, Smolec, Schwab, Gautschy, Bildsten,
  Cantiello, Dotter, Farmer, Goldberg, Jermyn, Kanbur, Marchant, Thoul,
  Townsend, Wolf, Zhang, \& Timmes}]{Paxton:2019aa}
Paxton, B., Smolec, R., Schwab, J., Gautschy, A., Bildsten, L., Cantiello, M.,
  Dotter, A., Farmer, R., Goldberg, J., Jermyn, A., Kanbur, S., Marchant, P.,
  Thoul, A., Townsend, R., Wolf, W., Zhang, M., \& Timmes, F. 2019, The
  Astrophysical Journal Supplement Series, 243, 10

\bibitem[{Perez-Becker \& Showman(2013)}]{Perez-Becker:2013fv}
Perez-Becker, D. \& Showman, A. 2013, The Astrophysical Journal, 776, 134

\bibitem[{Perna {et~al.}(2010)Perna, Menou, \& Rauscher}]{Perna_2010_2}
Perna, R., Menou, K., \& Rauscher, E. 2010, The Astrophysical Journal, 724, 313

\bibitem[{Rauscher(2017)}]{Rauscher:2017aa}
Rauscher, E. 2017, The Astrophysical Journal, 846, 69

\bibitem[{Rauscher \& Menou(2013)}]{Rauscher_2013}
Rauscher, E. \& Menou, K. 2013, The Astrophysical Journal, 764, 103

\bibitem[{Rauscher \& Showman(2014)}]{rauscher_showman_2013}
Rauscher, E. \& Showman, A. 2014, The Astrophysical Journal, 784, 160

\bibitem[{Rogers \& Komacek(2014)}]{Rogers:2014}
Rogers, T. \& Komacek, T. 2014, The Astrophysical Journal, 794, 132

\bibitem[{Rogers \& Showman(2014)}]{Rogers:2020}
Rogers, T. \& Showman, A. 2014, The Astrophysical Journal Letters, 782, L4

\bibitem[{Sainsbury-Martinez {et~al.}(2019)Sainsbury-Martinez, Wang, Fromang,
  Tremblin, Dubos, Meurdesoif, Spiga, Leconte, Baraffe, Chabrier, Mayne,
  Drummond, \& Debras}]{Sainsbury-Martinez:2019aa}
Sainsbury-Martinez, F., Wang, P., Fromang, S., Tremblin, P., Dubos, T.,
  Meurdesoif, Y., Spiga, A., Leconte, J., Baraffe, I., Chabrier, G., Mayne, N.,
  Drummond, B., \& Debras, F. 2019, Astronomy {\&} Astrophysics, 632

\bibitem[{Saumon {et~al.}(1995)Saumon, Chabrier, \& Horn}]{Saumon:1995}
Saumon, D., Chabrier, G., \& Horn, H.~V. 1995, The Astrophysical Journal
  Supplement Series, 99, 713

\bibitem[{Showman \& Guillot(2002)}]{showman_2002}
Showman, A. \& Guillot, T. 2002, Astronomy and Astrophysics, 385, 166

\bibitem[{Showman {et~al.}(2015)Showman, Lewis, \& Fortney}]{Showman:2014}
Showman, A., Lewis, N., \& Fortney, J. 2015, The Astrophysical Journal, 801, 95

\bibitem[{Spiegel \& Burrows(2013)}]{Spiegel:2013}
Spiegel, D. \& Burrows, A. 2013, The Astrophysical Journal, 772, 76

\bibitem[{Thorngren \& Fortney(2018)}]{Thorngren:2017}
Thorngren, D. \& Fortney, J. 2018, The Astronomical Journal, 155, 214

\bibitem[{Thorngren {et~al.}(2019)Thorngren, Gao, \&
  Fortney}]{Thorngren:2019aa}
Thorngren, D., Gao, P., \& Fortney, J. 2019, The Astrophysical Journal Letters,
  884, L6

\bibitem[{{Thorngren et al.}(2020)}]{Thorngren-et-al.:2019aa}
{Thorngren et al.} 2020, In preparation

\bibitem[{Tremblin {et~al.}(2017)Tremblin, Chabrier, Mayne, Amundsen, Baraffe,
  Debras, Drummond, Manners, \& Fromang}]{Tremblin:2017}
Tremblin, P., Chabrier, G., Mayne, N., Amundsen, D., Baraffe, I., Debras, F.,
  Drummond, B., Manners, J., \& Fromang, S. 2017, The Astrophysical Journal,
  843, 30

\bibitem[{Valsecchi {et~al.}(2015)Valsecchi, Rappaport, Rasio, Marchant, \&
  Rogers}]{Valsecchi:2015}
Valsecchi, F., Rappaport, S., Rasio, F., Marchant, P., \& Rogers, L. 2015, The
  Astrophysical Journal, 813, 101

\bibitem[{Weiss {et~al.}(2013)Weiss, Marcy, Rowe, Howard, Isaacson, Fortney,
  Miller, Demory, Fischer, Adams, Dupree, Howell, Kolbl, Johnson, Horch,
  Everett, Fabrycky, \& Seager}]{Weiss:2013}
Weiss, L., Marcy, G., Rowe, J., Howard, A., Isaacson, H., Fortney, J., Miller,
  N., Demory, B., Fischer, D., Adams, E., Dupree, A., Howell, S., Kolbl, R.,
  Johnson, J., Horch, E., Everett, M., Fabrycky, D., \& Seager, S. 2013, The
  Astrophysical Journal, 768, 14

\bibitem[{Wu \& Lithwick(2013)}]{Wu:2013}
Wu, Y. \& Lithwick, Y. 2013, The Astrophysical Journal, 763, 13

\bibitem[{Youdin \& Mitchell(2010)}]{Youdin_2010}
Youdin, A. \& Mitchell, J. 2010, The Astrophysical Journal, 721, 1113

\bibitem[{Zhang \& Showman(2018)}]{Zhang:2018aa}
Zhang, X. \& Showman, A. 2018, The Astrophysical Journal, 866, 2

\end{thebibliography}


\begin{thebibliography}
\end{thebibliography}
\fi

\if\bibinc y

\fi

\end{document}